\documentclass[11pt]{article}

\usepackage{amssymb}
\usepackage{url}
\usepackage{color}
\usepackage[a4paper,text={16.0cm,23.0cm},centering]{geometry}
\usepackage[figuresright]{rotating}
\usepackage{appendix}
\usepackage{graphicx}
\usepackage{epstopdf, epsfig}
\usepackage{bm}
\usepackage{amssymb}
\usepackage{amsmath}
\usepackage{mathrsfs}
\usepackage{amsfonts}
\usepackage{mathtools}
\usepackage{color}

\def\rd{{\rm d}}
\def\fr {\mbox{$\frac{1}{2}$}}

\def\qand{\quad\mbox{and}\quad}


\def\rd{{\rm d}}
\def\fr {\mbox{$\frac{1}{2}$}}

\newcounter{qcounter}

\newcommand{\squishlist}{
 \begin{list}{\bf--\arabic{qcounter}--~}{\usecounter{qcounter}}
  { \setlength{\itemsep}{0pt}
     \setlength{\parsep}{3pt}
     \setlength{\topsep}{3pt}
     \setlength{\partopsep}{0pt}
     \setlength{\leftmargin}{1.5em}
     \setlength{\labelwidth}{1em}
     \setlength{\labelsep}{0.25em} } }

\newcommand{\squishend}{
  \end{list}  }

\begin{document}

\numberwithin{equation}{section}
\setcounter{equation}{0}

\begin{center}
\textbf{AN ALTERNATIVE VIEW ON THE BATEMAN-LUKE}\\[2mm]
\textbf{VARIATIONAL PRINCIPLE}\\
\vspace{0.5cm}
\textbf{Hamid Alemi Ardakani}\footnote[1]{\textsf{Email address for correspondence: h.alemi-ardakani@exeter.ac.uk}}\\
\vspace{0.2cm}
\small{Department of Mathematics, University of Exeter, Penryn Campus, Cornwall TR10 9FE, UK}\\[2mm]
\today
\end{center}
\vspace{0.2cm}

\noindent\textbf{Abstract.} A new derivation of the Bernoulli equation for water waves in three-dimensional rotating and translating coordinate systems is given. 
An alternative view on the Bateman-Luke variational principle is presented. 
The variational principle recovers the boundary value problem governing the motion 
of \emph{potential water waves} in a container undergoing prescribed rigid-body motion in three dimensions. A mathematical theory is 
presented for the problem of three-dimensional interactions between \emph{potential surface waves} and a floating structure with interior \emph{potential fluid sloshing}. 
The complete set of equations of motion for the exterior gravity-driven water waves, 
and the exact nonlinear hydrodynamic equations of motion for the linear momentum and angular momentum of 
the floating structure containing fluid, are derived from a second variational principle. 
The two-dimensional form of the 3--D variational principles and their corresponding partial differential equations are presented. 

\section{Introduction}
\label{sec-Introduction}

The Bateman-Luke variational principle (Bateman 1932; Luke 1967) for the problem of fluid sloshing in a container undergoing prescribed rigid-body 
motion in three dimensions is given by Lukovsky (1990), Lukovsky (2015), Faltinsen \& Timokha (2009), 
Timokha (2016) and Faltinsen \emph{et al.} (2000) as 
\begin{equation}\label{Bateman-Luke-VP}
\begin{array}{rcl}
&&\displaystyle\delta\mathcal{L}\left(\Phi,\xi\right) = \delta\int_{t_1}^{t_2}\int_{\mathcal{Q}\left(t\right)} 
-\rho\left(\frac{\partial\Phi}{\partial t}+\fr\nabla\Phi\cdot\nabla\Phi
-\nabla\Phi\cdot\left(\bm v_0+\bm\omega\times\bm r\right)+U\right) \rd\mathcal{Q}\,\rd t = 0 \,,
\end{array} 
\end{equation}
where $\mathcal{Q}\left(t\right)$ is the fluid volume bounded by the free surface $\Sigma\left(t\right)$ 
and the wetted tank surface $S\left(t\right)$, $\Phi\left(x,y,z,t\right)$ is the velocity potential of the 
interior irrotational flow in a moving coordinate system $Oxyz$ fixed with respect to the rigid tank, the 
origin of the moving coordinate system $Oxyz$ is in the unperturbed free surface and moves with the velocity 
$\bm v_0$ relative to a fixed coordinate system $O^\prime x^\prime y^\prime z^\prime$, $\xi\left(x,y,t\right)$ is the free surface height 
relative to the moving frame $O$, $\bm\omega$ is the angular velocity 
of the tank relative to the fixed coordinate system $O^\prime x^\prime y^\prime z^\prime$, and $U\left(x,y,z,t\right)$ is the 
gravity field potential defined as 
\begin{equation}\label{gravity-field-potential}
U\left(x,y,z,t\right) = -\bm g\cdot\bm r^\prime \quad\mbox{with}\quad 
\bm r^\prime = \bm r_0^\prime+\bm r \,,
\end{equation}
where $\bm r^\prime$ is the radius-vector of a point of the fluid-body system with respect to the fixed frame $O^\prime$, 
$\bm r_0^\prime$ is the radius-vector of the origin of the moving frame $O$ with respect to the origin of the fixed frame 
$O^\prime$, $\bm r$ is the radius-vector with respect to $O$ and $\bm g$ is the gravity acceleration vector. Taking the variations 
$\delta\Phi$ and $\delta\xi$ in the variational principle (\ref{Bateman-Luke-VP}) subject to the restrictions $\delta\Phi=0$ at the 
end points of the time interval, $t_1$ and $t_2$, gives the following boundary value problem (Faltinsen \emph{et al.} 2000) 
\begin{equation}\label{boundary-value-problem}
\left.
\begin{array}{rcl}
&&\displaystyle\Delta\,\Phi := \Phi_{xx}+\Phi_{yy}+\Phi_{zz} = 0 \quad\mbox{in}\quad \mathcal{Q}\left(t\right)\,, \\[3mm]
&&\displaystyle\frac{\partial\Phi}{\partial\bm n} = \bm v_0\cdot\bm n+\bm\omega\cdot\left(\bm r\times\bm n\right) 
\quad\mbox{on}\quad S\left(t\right) \,,\\[3mm]
&&\displaystyle\frac{\partial\Phi}{\partial\bm n} = \bm v_0\cdot\bm n+\bm\omega\cdot\left(\bm r\times\bm n\right)
+\frac{\xi_t}{\sqrt{1+\xi_x^2+\xi_y^2}} \quad\mbox{on}\quad \Sigma\left(t\right) \,,\\[3mm]
&&\displaystyle\frac{\partial\Phi}{\partial t}+\fr\nabla\Phi\cdot\nabla\Phi-\nabla\Phi\cdot\left(\bm v_0
+\bm\omega\times\bm r\right)+U = 0 \quad\mbox{on}\quad \Sigma\left(t\right) \,,
\end{array}
\right\}
\end{equation}
where $\bm n$ is the outer normal to the boundary of $\mathcal{Q}\left(t\right)$. The second equation in (\ref{boundary-value-problem}) 
gives the rigid-wall boundary condition, the third equation in (\ref{boundary-value-problem}) is the kinematic free surface boundary condition, and 
the last equation in (\ref{boundary-value-problem}) is the dynamic free surface boundary condition deduced from Bernoulli's equation. 
The Bernoulli equation for the hydrodynamic pressure $p$ in $\mathcal{Q}\left(t\right)$ 
takes the form (Faltinsen \emph{et al.} 2000; Faltinsen \& Timokha 2009) 
\begin{equation}\label{Bernoulli-equation-1}
\frac{\partial\Phi}{\partial t}+\frac{p}{\rho}+\fr\nabla\Phi\cdot\nabla\Phi-\nabla\Phi\cdot\left(\bm v_0
+\bm\omega\times\bm r\right)+U = 0 \,,
\end{equation}
where $\rho$ is the density of the fluid and $\partial\Phi/\partial t$ is calculated in the moving coordinate system, i.e. for a point 
rigidly connected with the system $Oxyz$. 

It is stated in the literature that Bernoulli's equation which is a result of integrating 
the Euler equations relative to the fixed coordinate system, i.e. $O^\prime x^\prime y^\prime z^\prime$, is only valid in an inertial system 
and hence cannot directly be applied to an accelerated coordinate system, i.e. $Oxyz$. Hence, the Bernoulli equation (\ref{Bernoulli-equation-1}), 
in the moving coordinate system, is obtained by transforming the Bernoulli equation from the fixed coordinate system $O^\prime$ 
to the moving coordinate system $O$, by relating $\partial\Phi/\partial t$ between the inertial and moving coordinate systems. 

Our main goal in the first part of the current paper, \S\ref{sec-Bernoulli-equation} and \S\ref{sec-Bateman-Luke-variational-principle}, 
is to present a new derivation of the Bernoulli equation (\ref{Bernoulli-equation-1}) by integrating the Euler equations relative to the rotating and translating coordinate system 
attached to the moving container, using the vorticity equation. 
The proposed Bernoulli equation is then used to present an alternative view on the Bateman-Luke variational principle (\ref{Bateman-Luke-VP}) 
and the boundary value problem (\ref{boundary-value-problem}) for water waves in moving coordinate systems. 

Variational principles for the motion of a rigid body dynamically coupled to its interior fluid motion are given by Moiseyev \& Rumyantsev (1968) and 
Lukovsky (2015) (and references therein). In the work by Lukovsky (2015), the Bateman-Luke variational principle is 
used to develop a mathematical theory for interactions between potential surface waves and a floating rigid body containing cavities filled partially 
with a homogeneous incompressible ideal liquid. Alemi Ardakani (2019) derived a variational principle for the three-dimensional interactions between gravity-driven 
potential water waves and a floating rigid body dynamically coupled to its interior inviscid and incompressible fluid sloshing governed by the Euler equations 
relative to the rotating-translating coordinate system attached to the body. 
The variational principle gives the complete set of equations of motion for the exterior water waves, 
the Euler-Poincar\'{e} equations for the angular momentum and linear momentum of the rigid-body, and the Euler equations 
for the motion of the interior fluid of the rigid-body relative to the body coordinate system. 

The main goal in the second part of the current paper, \S\ref{sec-wave-structure-slosh-potential}, 
is to develop a mathematical theory for three-dimensional (3--D) interactions between \emph{potential surface waves} and a floating 
structure dynamically coupled to its interior \emph{potential fluid sloshing} relative to the rotating and translating coordinate 
system attached to the moving body. 
The Bateman-Luke variational principle, presented in \S\ref{sec-Bateman-Luke-variational-principle}, recovers the Neumann boundary value 
problem, in terms of the velocity potential, governing the motion of the interior fluid of the floating rigid-body interacting with the 
exterior ocean waves. The aim in \S\ref{sec-wave-structure-slosh-potential} is to present 
a second variational principle which recovers the equations of motion for the exterior potential water waves, and gives the exact hydrodynamic equations 
of motion for the angular momentum and linear momentum of the floating rigid-body dynamically coupled to its interior potential fluid motion. 
Adapting the variational principles developed by Alemi Ardakani (2019), the required variational principle takes the form 
\begin{figure}
\begin{center}
\includegraphics[width=14cm,height=4.7cm]{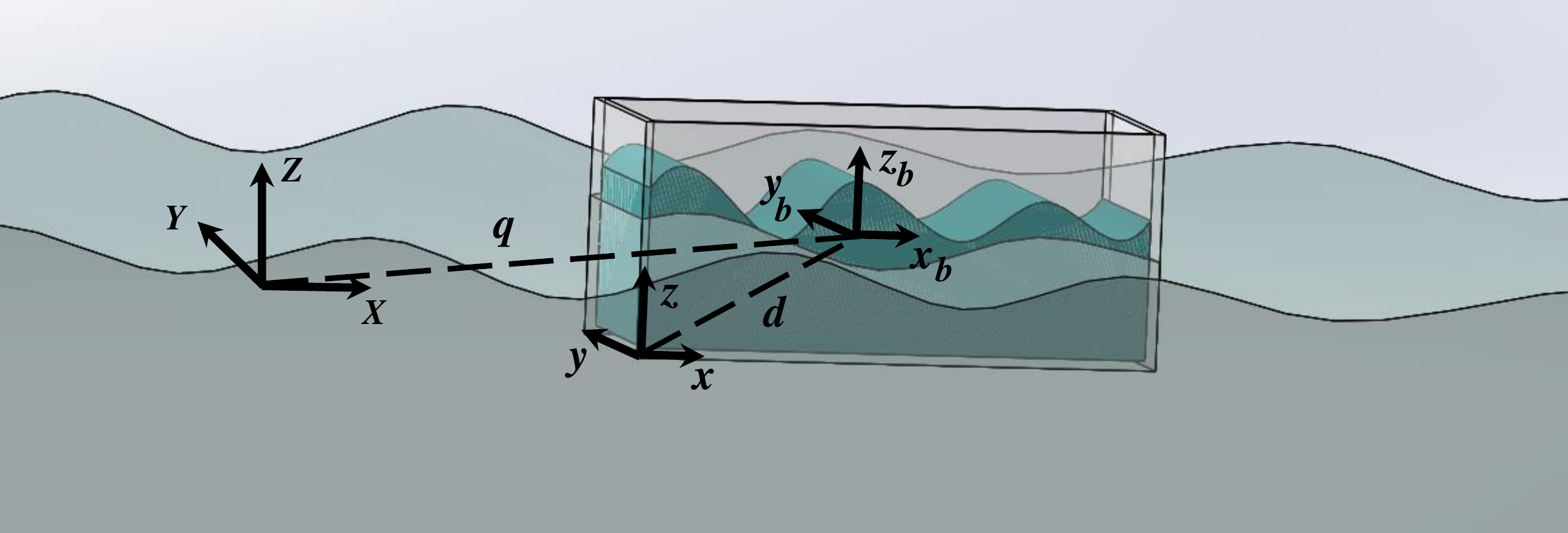}
\caption{Schematic showing a floating structure containing fluid in hydrodynamic interaction with exterior ocean waves.}
\label{fig-wave-structure-slosh}
\end{center}
\end{figure}
\begin{equation}\label{Lagrangian-wave-body-fluid-interaction}
\displaystyle\delta\mathscr{L}\left(\phi,\eta,\bm\Omega,\bm Q,\bm q,\dot{\bm q}\right) = 
\delta\mathscr{L}_1\left(\phi,\eta\right)+\delta\mathscr{L}_2\left(\bm\Omega,\bm Q,\bm q,\dot{\bm q}\right) = 0 \,,
\end{equation}
where 
\begin{equation}\label{L-1}
\begin{array}{rcl}
&&\displaystyle\mathscr{L}_1\left(\phi,\eta\right) = \displaystyle\int_{t_1}^{t_2}\int_{V\left(t\right)}
-\rho\left(\phi_t+\fr\nabla\phi\cdot\nabla\phi+gZ\right) \rd V\,\rd t \,,
\end{array}
\end{equation}
and 
\begin{equation}\label{L-2}
\begin{array}{rcl}
&&\displaystyle\mathscr{L}_2\left(\bm\Omega,\bm Q,\bm q,\dot{\bm q}\right) = \int^{t_2}_{t_1} \left(
\int_{\mathcal{Q}}\left(\fr\left\Arrowvert \bm u\right\Arrowvert^{2}
+\bm u\cdot\left(\bm\Omega\times\left(\bm x+\bm d\right)+\bm Q^T\dot{\bm q}\right)
+\bm Q^T\dot{\bm q}\cdot\left(\bm\Omega\times\left(\bm x+\bm d\right)\right)\right.\right. \\[3mm]
&&\displaystyle\left.\left.\hspace{5cm}+\fr\left\Arrowvert\dot{\bm q}\right\Arrowvert^{2}-g\left(\bm Q\left(\bm x+\bm d\right)
+\bm q\right)\cdot\bm k\right) \rho \,\rd\bm x 
+\fr\bm\Omega\cdot\bm I_f\bm\Omega \right. \\[3mm]
&&\displaystyle\left.\hspace{3cm}+\fr m_v\left\Arrowvert\dot{\bm q}\right\Arrowvert^{2}
+\left(\bm\Omega\times m_v\overline{\bm x}_v\right)\cdot\bm Q^T\dot{\bm q}
+\fr\bm\Omega\cdot\bm I_v\bm\Omega-m_vg\left(\bm Q\overline{\bm x}_v
+\bm q\right)\cdot\bm k \right)  \rd t \,,
\end{array}
\end{equation}
where in the derivation of the Lagrangian functional (\ref{Lagrangian-wave-body-fluid-interaction}), three frames of reference are used. 
The spatial frame, which is fixed in space, has coordinates denoted by $\bm X=\left(X,Y,Z\right)$. 
The first body frame, which is placed at the centre of rotation of the moving body 
and used for the analysis of the rigid body motion, has coordinates 
denoted by $\bm x_b=\left(x_b,y_b,z_b\right)$. The second body frame, 
which is attached to the moving body and used for the analysis of the fluid motion inside the tank, has coordinates denoted by $\bm x=\left(x,y,z\right)$. 
The distance between the origin of the body frame $\bm x$ to the point of rotation, i.e. the origin of the body frame $\bm x_b$, 
is denoted by $\bm d=\left(d_1,d_2,d_3\right)$ which is a constant vector. So the position of a fluid particle relative to the 
body frame $\bm x_b$ is $\bm x_b=\bm x+\bm d$. 
The fluid-tank system has a uniform translation $\bm q\left(t\right)=\left(q_1(t),q_2(t),q_3(t)\right)$ relative to the spatial frame $\bm X$, 
which is the vector from the origin of the spatial frame $\bm X$ to the origin of the body frame $\bm x_b$. In (\ref{Lagrangian-wave-body-fluid-interaction}), 
$\bm u\left(x,y,z,t\right)$ is the Eulerian velocity of a fluid particle in the body frame $\bm x$, $\mathcal{Q}$ is the volume of the fluid 
inside the tank, $\bm k$ is the unit vector in the $Z$ direction, $g$ is the acceleration due to gravity, 
$\rho$ is the water density which is assumed to be the same 
for the interior and exterior fluids, $\bm Q\left(t\right)\in \mbox{SO}\left(3\right)$ is a proper rotation in $\mathbb R^3$, i.e. 
$\bm Q^T\bm Q = \bm I$ and ${\rm det}\left(\bm Q\right)=1$, the \emph{body angular velocity} is a time-dependent vector 
$\bm\Omega\left(t\right) = \left(\Omega_1\left(t\right),\Omega_2\left(t\right),\Omega_3\left(t\right)\right)$ 
relative to the body coordinate system $\bm x_b$ with entries determined from the rotation tensor $\bm Q\left(t\right)$ by $\widehat{\bm\Omega}=\bm Q^T\dot{\bm Q}$ 
such that the skew-symmetric matrix $\widehat{\bm\Omega}$ satisfies 
$\widehat{\bm\Omega}\bm{\mathrm{r}}=\bm\Omega\times\bm{\mathrm{r}}$ for any $\bm{\mathrm{r}}\in\mathbb{R}^3$ (see Alemi Ardakani (2019) for 
more details), $\bm I_f$ which is defined as 
\begin{equation}\label{If}
\bm I_f = \int_{\mathcal{Q}}
\left(\left\Arrowvert\bm x+\bm d\right\Arrowvert^{2}\bm I-\left(\bm x+\bm d\right)\otimes
\left(\bm x+\bm d\right)\right)\rho \,\rd\bm{x} \,,
\end{equation}
is the mass moment of inertia of the interior fluid relative to the point of rotation, i.e. the origin of the body frame $\bm x_b$, 
$\otimes$ denotes the tensor product, $\bm I$ is the $3\times3$ identity matrix, 
$\bm I_v$ is the mass moment of inertia of the dry floating body relative to 
the point of rotation, $m_v$ is the mass of the dry body, $\overline{\bm x}_v=\left(\overline x_v,\overline y_v,\overline z_v\right)$ 
is the centre of mass of the dry body relative to the body frame $\bm x_b$, $\phi\left(X,Y,Z,t\right)$ and $\eta\left(X,Y,t\right)$ 
are respectively the velocity potential and the free surface elevation of the exterior irrotational water waves, 
and $V\left(t\right)$ is the transient domain of the exterior fluid. The configuration 
of the fluid in a rotating and translating floating structure in hydrodynamic interaction 
with exterior water waves is schematically shown in Figure \ref{fig-wave-structure-slosh}. 

The paper starts with the derivation of the Bernoulli equation and the Neumann boundary value problem for 
an inviscid and incompressible fluid sloshing in a container undergoing prescribed rigid-body motion in three dimensions 
in \S\ref{sec-Bernoulli-equation}. In \S\ref{sec-Bateman-Luke-variational-principle}, the Bateman-Luke variational principle 
is revisited for the problem of fluid sloshing in rotating and translating coordinates. In \S\ref{sec-wave-structure-slosh-potential}, 
a variational principle is given for 3--D interactions between ocean waves and a floating structure containing fluid. 
The paper ends with concluding remarks in \S\ref{sec-concluding-remarks}. 
In Appendix \ref{sec-2D-VP}, the proposed 3--D variational principles of 
\S\ref{sec-Bateman-Luke-variational-principle} and \S\ref{sec-wave-structure-slosh-potential} 
are reduced for the problem of two-dimensional (2--D) interactions between potential surface waves and a floating structure dynamically coupled 
to its interior potential fluid sloshing. The 2--D wave--body--slosh variational principles are given in Alemi Ardakani (2017). However, the presented 
variational principles of Alemi Ardakani (2017) are modified in Appendix \ref{sec-2D-VP}. 

\section{Derivation of the Bernoulli equation for the pressure field in rotating and translating coordinate systems}
\label{sec-Bernoulli-equation}

The configuration of the fluid in a rotating-translating rectangular vessel is schematically shown in Figure \ref{fig-wave-structure-slosh}. 
The vessel is a rigid body, which is free to rotate or translate in $\mathbb{R}^3$. The vessel is partially filled with 
an inviscid and incompressible fluid. The position of a fluid particle in the body frame $\bm x$ is related 
to a point in the spatial frame $\bm X$ by 
\begin{equation}\label{particle-position-x-X}
\bm X = \bm Q\left(\bm x+\bm d\right)+\bm q \,, 
\end{equation}
and the Eulerian velocity of a fluid particle $\bm u\left(\bm x,t\right)$ in the body frame $\bm x$ is related to the 
its velocity $\bm U\left(\bm X,t\right)$ in the spatial frame $\bm X$ by (Alemi Ardakani \& Bridges 2011; Alemi Ardakani 2019) 
\begin{equation}\label{particle-velocity-x-X}
\bm U = \bm Q\left(\bm u+\bm\Omega\times\left(\bm x+\bm d\right)+\bm Q^T\dot{\bm q}\right) \,,
\end{equation}
where the rotation tensor $\bm Q$ and the body angular velocity $\bm\Omega$ are defined in \S\ref{sec-Introduction}. 
It is shown in Alemi Ardakani \& Bridges (2011) that the Euler equations for the motion of an inviscid and incompressible 
fluid relative to the body coordinate system $\bm x$ takes the form 
\begin{equation}\label{Euler-equations}
\frac{D\bm u}{Dt}+\frac{1}{\rho}\nabla p = 
-2\bm\Omega\times\bm u-\dot{\bm\Omega}\times\left(\bm x+\bm d\right)
-\bm\Omega\times\left(\bm\Omega\times\left(\bm x+\bm d\right)\right)-\bm Q^T\bm g-\bm Q^T\ddot{\bm q} \,,
\end{equation}
where $\bm u=\left(u,v,w\right)$, $D\bm u/Dt=\bm u_t+\left(\bm u\cdot\nabla\right)\bm u$ and $\bm g=g\bm k$. The term 
$\bm Q^T\bm g$ rotates the usual gravity vector so that its direction is viewed properly in the body frame $\bm x_b$. 
The same is true for the translational acceleration $\ddot{\bm q}$. 

The fluid occupies the region $\mathcal{Q}\left(t\right)$ in the vessel which is bounded by the free surface 
$\Sigma\left(t\right)$ and the wetted tank surface $S\left(t\right)$, 
\begin{equation}\label{fluid-region}
0\le x\le L_1 \,,\quad 0\le y\le L_2 \,,\quad 0\le z\le h\left(x,y,t\right) \,,
\end{equation}
where the lengths $L_1$ and $L_2$ are given positive constants, and $z=h\left(x,y,t\right)$ is the 
position of the free surface inside the vessel, relative to the body frame $\bm x$. 

Conservation of mass relative to the body frame $\bm x$ takes the form 
\begin{equation}\label{mass-Conservation}
\nabla\cdot\bm u = u_x+v_y+w_z = 0 \,.
\end{equation}
The boundary conditions are 
\begin{equation}\label{no-flow-boundary-conditions}
\left.
\begin{array}{rcl}
u &=& 0 \quad\mbox{at}\quad x=0 \qand x=L_1 \,,\\[2mm]
v &=& 0 \quad\mbox{at}\quad y=0 \qand y=L_2 \,,\\[2mm]
w &=& 0 \quad\mbox{at}\quad z=0 \,,
\end{array}
\right\}
\end{equation}
which are the no-flow boundary conditions on the rigid walls, and at the free surface, 
the kinematic and dynamic boundary conditions are respectively 
\begin{equation}\label{free-surface-bc}
w = h_t+uh_x+vh_y \qand p = 0 \quad\mbox{at}\quad z=h\left(x,y,t\right) \,,
\end{equation}
where the surface tension is neglected in the boundary condition for the pressure $p$. 

The vorticity vector is defined by 
\begin{equation}\label{vorticity-vector}
\bm{\mathcal{V}} := \nabla\times\bm u \,. 
\end{equation}
Differentiating this equations gives 
\begin{equation}\label{vorticity-equation}
\frac{D\bm{\mathcal{V}}}{Dt} = \bm{\mathcal{V}}\cdot\nabla\bm u+\nabla\times\left(\frac{D\bm u}{Dt}\right) \,. 
\end{equation}
Taking the curl of the Euler equations (\ref{Euler-equations}) gives 
\begin{equation}\label{curl-of-Euler}
\nabla\times\left(\frac{D\bm u}{Dt}\right) = 2\bm\Omega\cdot\nabla\bm u-2\dot{\bm\Omega} \,.
\end{equation}
Substitution of (\ref{curl-of-Euler}) into (\ref{vorticity-equation}) gives the vorticity equation 
\begin{equation}\label{vorticity-differential-equation}
\frac{D\bm{\mathcal{V}}}{Dt} = \left(2\bm\Omega+\bm{\mathcal{V}}\right)\cdot\nabla\bm u-2\dot{\bm\Omega} \,.
\end{equation}
Now, if we set 
\begin{equation}\label{vorticity}
\bm{\mathcal{V}} = -2\bm\Omega \,,
\end{equation}
then 
\begin{equation}\label{vorticity-dot}
\frac{D\bm{\mathcal{V}}}{Dt} = -2\dot{\bm\Omega} \,,
\end{equation}
and the vorticity equation (\ref{vorticity-differential-equation}) is satisfied. 
Equation (\ref{vorticity}) will be important in the derivation of Bernoulli's equation in rotating and translating coordinate systems. 

Using equation (\ref{vorticity}), the vector identity 
\begin{equation}\label{vector-identity}
\frac{D\bm u}{Dt} = \frac{\partial\bm u}{\partial t}+\nabla\left(\fr\bm u\cdot\bm u\right)-\bm u\times\left(\nabla\times\bm u\right) \,,
\end{equation}
takes the form 
\begin{equation}\label{vector-identity-with-vorticity}
\frac{D\bm u}{Dt} = \frac{\partial\bm u}{\partial t}+\nabla\left(\fr\bm u\cdot\bm u\right)-2\bm\Omega\times\bm u \,.
\end{equation}
Using the vector identity (\ref{vector-identity-with-vorticity}) the Euler equations (\ref{Euler-equations}) reduces to 
\begin{equation}\label{Euler-equations-reduced}
\frac{\partial\bm u}{\partial t}+\nabla\left(\fr\bm u\cdot\bm u\right)
+\frac{1}{\rho}\nabla p+\dot{\bm\Omega}\times\left(\bm x+\bm d\right)
+\bm\Omega\times\left(\bm\Omega\times\left(\bm x+\bm d\right)\right)+\bm Q^T\bm g+\bm Q^T\ddot{\bm q} = \bm 0 \,.
\end{equation}
Now, if we introduce a velocity potential $\Phi\left(x,y,z,t\right)$ such that 
\begin{equation}\label{velocity-field}
\bm u\left(\bm x,t\right) = \nabla\Phi\left(\bm x,t\right)-\bm\Omega\times\left(\bm x+\bm d\right)
-\bm Q^T\dot{\bm q} \,,
\end{equation}
then the velocity field in (\ref{velocity-field}) satisfies the vorticity equation. Noting that 
(Marsden \& Ratiu 1999; Holm, Schmah \&  Stoica 2009) 
\begin{equation}\label{Q-t}
\frac{d}{dt}\left(\bm Q^{-1}\right) = -\bm Q^{-1}\dot{\bm Q}\bm Q^{-1}  
\end{equation}
then 
\begin{equation}\label{u-t}
\frac{\partial\bm u}{\partial t} = \nabla\Phi_t-\dot{\bm\Omega}\times\left(\bm x+\bm d\right)+\bm\Omega\times\bm Q^T\dot{\bm q}
-\bm Q^T\ddot{\bm q} \,.
\end{equation}
Substitution of (\ref{velocity-field}) and (\ref{u-t}) into the Euler equations (\ref{Euler-equations-reduced}) gives 
\begin{equation}\label{Euler-equations-reduced-1}
\left.
\begin{array}{rcl}
&&\displaystyle\nabla\Phi_t+\bm\Omega\times\left(\bm\Omega\times\left(\bm x+\bm d\right)+\bm Q^T\dot{\bm q}\right)
+\frac{1}{\rho}\nabla p+\bm Q^T\bm g \\[2mm]
&&\displaystyle+\nabla\left[\fr\nabla\Phi\cdot\nabla\Phi
-\nabla\Phi\cdot\left(\bm\Omega\times\left(\bm x+\bm d\right)+\bm Q^T\dot{\bm q}\right)\right. \\[2mm]
&&\displaystyle\left.\hspace{7mm}+\fr\bm\Omega\times\left(\bm x+\bm d\right)\cdot\bm\Omega\times\left(\bm x+\bm d\right)
+\bm\Omega\times\left(\bm x+\bm d\right)\cdot\bm Q^T\dot{\bm q}\right] = \bm 0 \,.
\end{array}
\right\}
\end{equation}
But 
\begin{equation}\label{simplification-1}
\nabla\left(\bm\Omega\times\left(\bm x+\bm d\right)\cdot\bm Q^T\dot{\bm q}\right) = 
-\bm\Omega\times\bm Q^T\dot{\bm q} \,,
\end{equation}
and 
\begin{equation}\label{simplification-2}
\nabla\left(\fr\bm\Omega\times\left(\bm x+\bm d\right)\cdot\bm\Omega\times\left(\bm x+\bm d\right)\right) = 
-\bm\Omega\times\left(\bm\Omega\times\left(\bm x+\bm d\right)\right) \,,
\end{equation}
and so equation (\ref{Euler-equations-reduced-1}) simplifies to 
\begin{equation}\label{Eule-equations-reduced-2}
\nabla\left(\Phi_t+\frac{p}{\rho}+\fr\nabla\Phi\cdot\nabla\Phi
-\nabla\Phi\cdot\left(\bm\Omega\times\left(\bm x+\bm d\right)+\bm Q^T\dot{\bm q}\right)
+\bm Q^T\bm g\cdot\left(\bm x+\bm d\right)\right) = \bm 0 \,,
\end{equation}
or 
\begin{equation}\label{Bernoulli-equation}
\Phi_t+\frac{p}{\rho}+\fr\nabla\Phi\cdot\nabla\Phi
-\nabla\Phi\cdot\left(\bm\Omega\times\left(\bm x+\bm d\right)+\bm Q^T\dot{\bm q}\right)
+\bm Q^T\bm g\cdot\left(\bm x+\bm d\right) = Be\left(t\right) \,,
\end{equation}
where $Be\left(t\right)$ is the Bernoulli function which can be absorbed into $\Phi\left(\bm x,t\right)$. Therefore, 
Bernoulli's equation for the pressure field $p$ in $\mathcal{Q}\left(t\right)$ takes the form 
\begin{equation}\label{Bernoulli-equation-final}
\Phi_t+\frac{p}{\rho}+\fr\nabla\Phi\cdot\nabla\Phi
-\nabla\Phi\cdot\left(\bm\Omega\times\left(\bm x+\bm d\right)+\bm Q^T\dot{\bm q}\right)
+\bm Q^T\bm g\cdot\left(\bm x+\bm d\right) = 0 \,.
\end{equation}
All terms in the new Bernoulli equation (\ref{Bernoulli-equation-final}) are relative to the body frame attached to the 
rotating and translating rigid body. 

In terms of the velocity potential $\Phi\left(x,y,z,t\right)$, the rigid-wall boundary conditions in (\ref{no-flow-boundary-conditions}) 
become 
\begin{equation}\label{no-flow-boundary-conditions-Phi}
\begin{array}{rcl}
&&\displaystyle\nabla\Phi = \bm\Omega\times\left(\bm x+\bm d\right)+\bm Q^T\dot{\bm q} \,\quad\mbox{on}\quad S\left(t\right) \,,
\end{array}
\end{equation}
which can be written in the form 
\begin{equation}\label{no-flow-boundary-conditions-Phi-1}
\begin{array}{rcl}
&&\displaystyle\frac{\partial\Phi}{\partial\bm n_b} 
= \bm\Omega\cdot\left(\left(\bm x+\bm d\right)\times\bm n_b\right)
+\bm Q^T\dot{\bm q}\cdot\bm n_b \,\quad\mbox{on}\quad S\left(t\right) \,,
\end{array}
\end{equation}
where $\bm n_b=\bm Q^{-1}\bm n$ is outer normal to the boundary of $\mathcal{Q}\left(t\right)$ relative to the body frame $\bm x_b$. 
Also in terms of the velocity potential $\Phi\left(x,y,z,t\right)$, the kinematic free surface boundary condition 
in (\ref{free-surface-bc}) becomes 
\begin{equation}\label{KFSBC-Phi}
\begin{array}{rcl}
&&\displaystyle\frac{\partial\Phi}{\partial\bm n_b} = \bm\Omega\cdot\left(\left(\bm x+\bm d\right)\times\bm n_b\right)
+\bm Q^T\dot{\bm q}\cdot\bm n_b+\frac{h_t}{\sqrt{1+h_x^2+h_y^2}} \,\quad\mbox{on}\quad \Sigma\left(t\right) \,.
\end{array}
\end{equation}
From (\ref{Bernoulli-equation-final}) it can be concluded that the dynamic free surface boundary condition in (\ref{free-surface-bc}) becomes 
\begin{equation}\label{DFSBC-Phi}
\Phi_t+\fr\nabla\Phi\cdot\nabla\Phi
-\nabla\Phi\cdot\left(\bm\Omega\times\left(\bm x+\bm d\right)+\bm Q^T\dot{\bm q}\right)
+\bm Q^T\bm g\cdot\left(\bm x+\bm d\right) = 0 \,\quad\mbox{on}\quad \Sigma\left(t\right) \,.
\end{equation}
Finally, substitution of the velocity field (\ref{velocity-field}) into the continuity equation (\ref{mass-Conservation}) 
leads to Laplace's equation for $\Phi\left(x,y,z,t\right)$, 
\begin{equation}\label{Laplace-equation}
\Delta\,\Phi = \Phi_{xx}+\Phi_{yy}+\Phi_{zz} = 0  \,\quad\mbox{in}\quad \mathcal{Q}\left(t\right) \,.
\end{equation}

In summary, the proposed derivations in this section lead to mathematically precise definitions for all terms 
in the Bernoulli equation (\ref{Bernoulli-equation-final}) and in the corresponding boundary value problem (\ref{no-flow-boundary-conditions-Phi-1}), 
(\ref{KFSBC-Phi}), (\ref{DFSBC-Phi}) and (\ref{Laplace-equation}) for water waves in moving coordinate systems. 

\section{The Bateman-Luke variational principle for fluid sloshing in vessels undergoing prescribed rigid-body 
motion in three dimensions}
\label{sec-Bateman-Luke-variational-principle}

Based on the new Bernoulli equation (\ref{Bernoulli-equation-final}), the Bateman-Luke variational 
principle (\ref{Bateman-Luke-VP}) is modified to 
\begin{equation}\label{pressure-integral}
\begin{cases}
\displaystyle\delta\mathscr{L}\left(\Phi,h\right) = \delta\int_{t_1}^{t_2} \int_{\mathcal{Q}\left(t\right)} 
p\left(\bm x,t\right) \,\rd\mathcal{Q}\,\rd t \\[3mm]
\displaystyle = \delta\int_{t_1}^{t_2}\int_{\mathcal{Q}\left(t\right)} 
-\left(\Phi_t+\fr\nabla\Phi\cdot\nabla\Phi
-\nabla\Phi\cdot\left(\bm\Omega\times\left(\bm x+\bm d\right)+\bm Q^T\dot{\bm q}\right)
+\bm Q^T\bm g\cdot\left(\bm x+\bm d\right)\right) \,\rho\,\rd\mathcal{Q}\,\rd t = 0 \,,
\end{cases}
\end{equation}
subject to the endpoint conditions $\delta\Phi\left(\bm x,t_1\right)=\delta\Phi\left(\bm x,t_2\right)=0$. 
Note that in (\ref{pressure-integral}), $\bm\Omega$ is the \emph{body} angular velocity, $\bm Q^T\dot{\bm q}$ 
is the translational velocity of the moving rigid body relative to the body coordinate system $\bm x_b$, and 
$\bm Q^T\bm g$ is relative to the body frame $\bm x_b$, while in (\ref{Bateman-Luke-VP}), $\bm\omega$ is the angular velocity 
of the rigid body relative to the spatial coordinate system, $\bm v_0$ is the translational velocity of the rigid body 
relative to the spatial frame, and $U=-\bm{g}\cdot\bm{r}^\prime$ is relative to the spatial frame. 

According to the usual procedure in the calculus of variations, the variational principle (\ref{pressure-integral}) 
becomes 
\begin{equation}\label{BL-1}
\left.
\begin{array}{rcl}
&&\displaystyle\delta\mathscr{L}\left(\Phi,h\right) = \int_{t_1}^{t_2} \int_{\Sigma\left(t\right)} 
p\left(x,y,h,t\right) \delta h\ell^{-1} \,\rd S\,\rd t \\[3mm]
&&\displaystyle+\int_{t_1}^{t_2}\int_{\mathcal{Q}\left(t\right)} 
-\left(\delta\Phi_t+\nabla\Phi\cdot\nabla\delta\Phi
-\nabla\delta\Phi\cdot\left(\bm\Omega\times\left(\bm x+\bm d\right)+\bm Q^T\dot{\bm q}\right)\right) 
\,\rho\,\rd\mathcal{Q}\,\rd t = 0 \,,
\end{array}
\right\}
\end{equation}
where $\ell=\sqrt{1+h_x^2+h_y^2}$ and 
\begin{equation}\label{pressure-free-surface}
\begin{array}{rcl}
p\left(x,y,h,t\right) &=& \displaystyle
-\rho\left(\Phi_t+\fr\nabla\Phi\cdot\nabla\Phi
-\nabla\Phi\cdot\left(\bm\Omega\times\left(\bm x+\bm d\right)+\bm Q^T\dot{\bm q}\right)
+\bm Q^T\bm g\cdot\left(\bm x+\bm d\right)\right)\bigg|^{z=h\left(x,y,t\right)} \,.
\end{array}
\end{equation}
But 
\begin{equation}\label{BL-2}
\int_{t_1}^{t_2}\int_{\mathcal{Q}\left(t\right)} -\delta\Phi_t \,\rho\,\rd\mathcal{Q}\,\rd t = 
\int_{t_1}^{t_2}\int_{\Sigma\left(t\right)} h_t\delta\Phi\ell^{-1} \,\rho\,\rd S\,\rd t \,,
\end{equation}
noting that $\delta\Phi=0$ at $t=t_1$ and $t=t_2$. Moreover, using Green's first identity we obtain 
\begin{equation}\label{BL-3}
\begin{cases}
\displaystyle-\int_{t_1}^{t_2}\int_{\mathcal{Q}\left(t\right)} \nabla\Phi\cdot\nabla\delta\Phi \,\rho\,\rd\mathcal{Q}\,\rd t 
= \int_{t_1}^{t_2}\int_{\mathcal{Q}\left(t\right)} \Delta\,\Phi\,\delta\Phi \,\rho\,\rd\mathcal{Q}\,\rd t
-\int_{t_1}^{t_2}\int_{\partial\mathcal{Q}} \nabla\Phi\cdot\bm n_b\,\delta\Phi\,\rho\,\rd S\,\rd t \\[3mm]
\displaystyle = \int_{t_1}^{t_2}\int_{\mathcal{Q}\left(t\right)}
\Delta\,\Phi\,\delta\Phi \,\rho\,\rd\mathcal{Q}\,\rd t
-\int_{t_1}^{t_2}\int_{S\left(t\right)} \nabla\Phi\cdot\bm n_b\,\delta\Phi\,\rho\,\rd S\,\rd t
-\int_{t_1}^{t_2}\int_{\Sigma\left(t\right)} \nabla\Phi\cdot\bm n_b\,\delta\Phi\,\rho\,\rd S\,\rd t \,,
\end{cases}
\end{equation}
where $\bm n_b$ is the unit outward normal vector to the boundary of $\mathcal{Q}\left(t\right)$, and also 
\begin{equation}\label{BL-4}
\left.
\begin{array}{rcl}
&&\displaystyle\int_{t_1}^{t_2}\int_{\mathcal{Q}\left(t\right)} 
\nabla\delta\Phi\cdot\left(\bm\Omega\times\left(\bm x+\bm d\right)
+\bm Q^T\dot{\bm q}\right) \,\rho\,\rd\mathcal{Q}\,\rd t \\[3mm]
&&\displaystyle = \int_{t_1}^{t_2}\int_{S\left(t\right)}
\left(\bm\Omega\times\left(\bm x+\bm d\right)+\bm Q^T\dot{\bm q}\right)\cdot\bm n_b\,\delta\Phi\,\rho\,\rd S\,\rd t \\[4mm]
&&\displaystyle+\int_{t_1}^{t_2}\int_{\Sigma\left(t\right)}
\left(\bm\Omega\times\left(\bm x+\bm d\right)+\bm Q^T\dot{\bm q}\right)\cdot\bm n_b\,\delta\Phi\,\rho\,\rd S\,\rd t \,.
\end{array}
\right\}
\end{equation}
Now, substitution of (\ref{BL-2}), (\ref{BL-3}) and (\ref{BL-4}) into the variational principle (\ref{BL-1}) gives 
\begin{equation}\label{BL-5}
\left.
\begin{array}{rcl}
&&\displaystyle\delta\mathscr{L}\left(\Phi,h\right) = \int_{t_1}^{t_2} \int_{\Sigma\left(t\right)} 
p\left(x,y,h,t\right) \delta h\ell^{-1} \,\rd S\,\rd t
+\int_{t_1}^{t_2}\int_{\mathcal{Q}\left(t\right)}
\Delta\,\Phi\,\delta\Phi \,\rho\,\rd\mathcal{Q}\,\rd t \\[3mm]
&&\displaystyle+\int_{t_1}^{t_2}\int_{\Sigma\left(t\right)} \left(h_t\ell^{-1}
+\left(-\nabla\Phi+\bm\Omega\times\left(\bm x+\bm d\right)+\bm Q^T\dot{\bm q}\right)\cdot\bm n_b\right)
\delta\Phi\,\rho\,\rd S\,\rd t \\[3mm]
&&\displaystyle+\int_{t_1}^{t_2}\int_{S\left(t\right)} \left(-\nabla\Phi
+\bm\Omega\times\left(\bm x+\bm d\right)+\bm Q^T\dot{\bm q}\right)\cdot\bm n_b\,\delta\Phi\,\rho\,\rd S\,\rd t = 0 \,.
\end{array}
\right\}
\end{equation}
From (\ref{BL-5}), it can be concluded that invariance of $\mathscr{L}$ with respect to a variation in the free surface 
height $h$ yields the dynamic free surface boundary condition (\ref{DFSBC-Phi}). Similarly, the invariance of $\mathscr{L}$ 
with respect to a variation in the velocity potential $\Phi$ at the free surface $\Sigma\left(t\right)$ yields the kinematic 
free surface boundary condition (\ref{KFSBC-Phi}), and the invariance of $\mathscr{L}$ with respect to a variation 
in the velocity potential $\Phi$ along the wetted surface $S\left(t\right)$ 
recovers the rigid-wall boundary conditions (\ref{no-flow-boundary-conditions-Phi-1}). Moreover, the invariance of $\mathscr{L}$ with 
respect to a variation in the velocity potential $\Phi$ yields the field equation (\ref{Laplace-equation}). 

\section{A variational principle for 3--D interactions between potential surface waves and a floating structure with 
interior potential fluid sloshing}
\label{sec-wave-structure-slosh-potential}

The interest in this section is to first present a variational principle for the 3--D rotational and translational motion 
of a rigid body containing fluid such that the interior fluid satisfies the velocity potential theory developed 
in \S\ref{sec-Bernoulli-equation} and \S\ref{sec-Bateman-Luke-variational-principle}, 
and then extend the variational principle for the problem of interactions between potential water waves 
and a floating structure dynamically coupled to its interior potential fluid sloshing. 

Gerrits \& Veldman (2003) and Veldman \emph{et al.} (2007) studied the problem of coupled liquid-solid dynamics 
for a liquid-filled spacecraft in three dimensions. They presented the differential equations for the motion of the spacecraft 
containing fluid, describing conservation of linear momentum and angular momentum. The equation for the 
conservation of angular momentum of the rigid body takes the form (see equation (6) of Gerrits and Veldman (2003) 
and the work of Alemi Ardakani (2019) for minor modifications of this equation) 
\begin{equation}\label{conservation-of-angular-momentum}
m_v\overline{\bm x}_v\times\bm Q^{-1}\ddot{\bm q}+\bm I_v\dot{\bm\Omega}
+\bm\Omega\times\bm I_v\bm\Omega = \bm{\mathcal{T}}-m_v g\overline{\bm x}_v\times\bm\chi \,,
\end{equation}
where $m_v$ is the mass of the dry body, $\overline{\bm x}_v$ is the centre of 
mass of the dry body relative to the body frame $\bm x_b$, $\bm I_v$ is the mass moment of inertia of the dry body relative to 
the point of rotation, i.e. the origin of the body frame $\bm x_b$, $\bm Q^{-1}\ddot{\bm q}$ is the linear acceleration of 
the origin of the moving coordinate frame relative to the body frame $\bm x_b$, and 
\begin{equation}\label{tau-definition}
\bm{\mathcal{T}} = \int_{\mathcal{Q}\left(t\right)} \left(\bm x+\bm d\right)\times\nabla p \,\rd\bm x \,,
\end{equation}
is the torque that the interior fluid exerts on the boundary of the rigid body via pressure, and 
\begin{equation}\label{chi-definition}
\bm\chi = \bm Q^{-1}\bm k \,,
\end{equation}
where $\bm k$ is the unit vector in the $Z$ direction. For our problem in this paper, the interior fluid is inviscid and 
incompressible and satisfies the velocity potential theory developed in \S\ref{sec-Bernoulli-equation} and \S\ref{sec-Bateman-Luke-variational-principle}. 
Now, after substituting for $\nabla p$ for the interior fluid of the rigid body 
from (\ref{Eule-equations-reduced-2}) in (\ref{tau-definition}), equation (\ref{conservation-of-angular-momentum}) 
for the body angular velocity $\bm\Omega\left(t\right)$ takes the form 
\begin{equation}\label{Omega-equation}
\begin{cases}
\displaystyle\int_{\mathcal{Q}} -\left(\bm x+\bm d\right)
\times\left[\nabla\Phi_t+\nabla\left(\fr\nabla\Phi\cdot\nabla\Phi\right)
-\nabla\left(\nabla\Phi\cdot\left(\bm\Omega\times\left(\bm x+\bm d\right)+\bm Q^T\dot{\bm q}\right)\right)\right] 
\rho\,\rd\bm x \\[3mm]
\displaystyle-mg\overline{\bm x}\times\bm\chi
-m_v\overline{\bm x}_v\times\bm Q^{-1}\ddot{\bm q}-\bm I_v\dot{\bm\Omega}
+\bm I_v\bm\Omega\times\bm\Omega = \bm 0 \,,
\end{cases}
\end{equation}
where $m$ is the total mass of the (interior fluid + body) system, 
\begin{equation}\label{total-mass}
m = m_f+m_v \quad\mbox{with}\quad m_f=\int_{\mathcal{Q}} \rho\,\rd\bm x \,,
\end{equation}
where $m_f$ is the mass of the interior fluid which is time independent. Note that in $(\ref{Omega-equation})$ 
$\overline{\bm x}$ is the centre of mass of the coupled (interior fluid + body) system relative to the body frame $\bm x_b$, 
which is time dependent and satisfies 
\begin{equation}\label{mx-total}
m\overline{\bm x}\left(t\right)=m_f\overline{\bm x}_f+m_v\overline{\bm x}_v \,,
\end{equation} 
with 
\begin{equation}\label{mf-xf}
m_f\overline{\bm x}_f = \int_{\mathcal{Q}} \left(\bm x+\bm d\right) \rho\,\rd\bm x \,,
\end{equation}
where $\overline{\bm x}_f\left(t\right)$ is the centre of mass of the interior fluid relative to the body frame $\bm x_b$. 

Similarly, the equation for the conservation of linear momentum of the rigid body takes the form (see equation (5) of Gerrits and Veldman (2003)) 
\begin{equation}\label{conservation-of-linear-momentum}
m_v\bm Q^{-1}\ddot{\bm q}+\dot{\bm\Omega}\times m_v\overline{\bm x}_v
+\bm\Omega\times\left(\bm\Omega\times m_v\overline{\bm x}_v\right) = \bm{\mathcal{F}}-m_v g\bm\chi \,,
\end{equation}
where 
\begin{equation}\label{F-definition}
\bm{\mathcal{F}} = \int_{\mathcal{Q}\left(t\right)} \nabla p \,\rd\bm x \,,
\end{equation}
is the force that the interior fluid exerts on the boundary of the rigid body via pressure. 
Now, after substituting for $\nabla p$ from (\ref{Eule-equations-reduced-2}) in (\ref{F-definition}), 
equation (\ref{conservation-of-linear-momentum}) for the translational motion of the rigid body $\bm q\left(t\right)$ takes the form 
\begin{equation}\label{q-equation}
\begin{cases}
\displaystyle\int_{\mathcal{Q}} 
-\left[\nabla\Phi_t+\nabla\left(\fr\nabla\Phi\cdot\nabla\Phi\right)
-\nabla\left(\nabla\Phi\cdot\left(\bm\Omega\times\left(\bm x+\bm d\right)+\bm Q^T\dot{\bm q}\right)\right)\right] 
\rho\,\rd\bm x \\[3mm]
\displaystyle-mg\bm\chi-m_v\bm Q^{-1}\ddot{\bm q}-\dot{\bm\Omega}\times m_v\overline{\bm x}_v
-\bm\Omega\times\left(\bm\Omega\times m_v\overline{\bm x}_v\right) = \bm 0 \,.
\end{cases}
\end{equation}

The Lagrangian action for the motion of a rigid-body dynamically coupled to its interior fluid motion takes the form 
\begin{equation} \label{Lag-action}
\mathscr{L} = \int^{t_{2}}_{t_{1}} \left(\mathrm{KE}^{fluid}-\mathrm{PE}^{fluid}+\mathrm{KE}^{body}
-\mathrm{PE}^{body}\right) \rd t \,,
\end{equation}
where 
$\mathrm{KE}^{fluid}$ is the kinetic energy of the fluid, $\mathrm{KE}^{body}$ is the kinetic energy of the rigid body, 
$\mathrm{PE}^{fluid}$ is the potential energy of the fluid and $\mathrm{PE}^{body}$ is the potential energy of the rigid body. 
It is shown in Alemi Ardakani (2019) that the action functional (\ref{Lag-action}) for a rigid body which contains 
an inviscid and incompressible fluid and undergoes three dimensional rotational and translational motions takes the form (\ref{L-2}). The equations of motion for 
the body angular velocity $\bm\Omega\left(t\right)$ and the translational motion $\bm q\left(t\right)$ 
of the rigid body are provided by Hamilton's variational principle:
\begin{equation}\label{Hamilton-principle}
\delta\mathscr{L}_2\left(\bm\Omega,\bm Q,\bm q,\dot{\bm q}\right) = 0 \,,
\end{equation}
subject to the fixed endpoints $\delta\bm q\left(t_1\right)=\delta\bm q\left(t_2\right)=\bm 0$, and noting that 
the variations $\delta\bm Q$ are taken among paths $\bm Q\left(t\right)\in \mbox{SO}\left(3\right)$, $t\in[t_1,t_2]$, with fixed endpoints, 
so that $\delta\bm Q\left(t_1\right)=\delta\bm Q\left(t_2\right)=\bm 0$. It is proved in Alemi Ardakani (2019) that taking the variations 
$\delta\bm\Omega$, $\delta\bm Q$, $\delta\bm q$ and $\delta\dot{\bm q}$ in the variational principle (\ref{Hamilton-principle}), 
using the Euler-Poincar\'{e} framework (Marsden \& Ratiu 1999; Holm, Schmah \&  Stoica 2009), gives the Euler-Poincar\'{e} equation 
for $\bm\Omega\left(t\right)$ as 
\begin{equation}\label{EP-Omega}
\begin{array}{rcl}
&&\displaystyle\int_{\mathcal{Q}} -\left(\bm x+\bm d\right)\times\left(\frac{D\bm u}{Dt}
+2\bm\Omega\times\bm u\right) \rho \,\rd\bm{x}-mg\overline{\bm x}\times\bm\chi
= m\overline{\bm x}\times\bm Q^{-1}\ddot{\bm q}
+\bm I_t\dot{\bm\Omega}+\bm\Omega\times\bm I_t\bm\Omega \,,
\end{array}
\end{equation}
and the Euler-Poincar\'{e} equation for $\bm q\left(t\right)$ as 
\begin{equation}\label{EP-q}
\begin{array}{rcl}
&&\displaystyle\int_{\mathcal{Q}} \left(-\frac{D\bm u}{Dt}-2\bm\Omega\times\bm u\right) \rho \,\rd\bm{x}
-m\bm Q^{-1}\ddot{\bm q}-\dot{\bm\Omega}\times m\overline{\bm x}
-\bm\Omega\times\left(\bm\Omega\times m\overline{\bm x}\right)-mg\bm\chi = \bm0 \,,
\end{array}
\end{equation}  
where $\bm I_t=\bm I_f+\bm I_v$ is the mass moment of inertia of the coupled (interior fluid + body) system. 

The interior fluid of the rigid body has a velocity field of the form (\ref{velocity-field}). 
Now, it can be proved that substitution of the velocity field (\ref{velocity-field}) into the Euler-Poincar\'{e} equation (\ref{EP-Omega}) recovers 
the $\bm\Omega$-equation (\ref{Omega-equation}) obtained from balance of angular momentum of the rigid body. Similarly, substitution of 
the velocity field (\ref{velocity-field}) into the Euler-Poincar\'{e} equation (\ref{EP-q}) recovers 
the $\bm q$-equation (\ref{q-equation}) obtained from balance of linear momentum of the rigid body. 

The variational principle (\ref{Hamilton-principle}), with the Lagrangian action defined in (\ref{L-2}), can be extended 
to the problem of 3--D water waves in hydrodynamic interaction with a freely floating rigid body 
containing fluid by the addition of Luke's variational principle (Luke 1967; Van Daalen, Van Groesen \& Zandbergen 1993; Alemi Ardakani 2019) 
to Hamilton's variational principle (\ref{Hamilton-principle}). The variational principle for the motion of the exterior water waves 
and the motion of the rigid body containing fluid takes the form (\ref{Lagrangian-wave-body-fluid-interaction}). See the work of 
Alemi Ardakani (2019) for more details. In (\ref{Lagrangian-wave-body-fluid-interaction}), 
$\phi\left(\bm X,t\right)$ is the velocity potential of the exterior irrotational fluid lying between $Z=-H\left(X,Y\right)$ 
and $Z=\eta\left(X,Y,t\right)$ with the gravity acceleration $g$ acting in the negative $Z$ direction.
In the horizontal directions $X$ and $Y$, the fluid domain is cut off by a cylindrical vertical surface $\mathcal{S}$ 
of infinite radius which extends from the bottom to the free surface, and the transient fluid domain 
$V\left(t\right)$ cosists of a fluid bounded by the impermeable bottom $S_b$ defined by the equation $Z=-H\left(X,Y\right)$, 
the free surface $S_\eta$ defined by the equation $Z=\eta\left(X,Y,t\right)$, the vertical surface $\mathcal{S}$ and the wetted surface 
$S_w$ of the rigid body interacting with exterior water waves. It can be proved that taking the variations $\delta\eta$ and 
$\delta\phi$ of the first component of the variational principle (\ref{Lagrangian-wave-body-fluid-interaction}) subject 
to the restrictions $\delta\phi=0$ at the end points of the time interval, $t_1$ and $t_2$, recovers the 
complete set of equations of motion for the classical water-wave problem in three dimensions as 
(Luke 1967; Miles 1977; Lukovsky 2015; Van Daalen, Van Groesen \& Zandbergen 1993) 
\begin{equation}\label{exterior-equations}
\left.
\begin{array}{rcl}
&&\displaystyle \Delta\,\phi := \phi_{XX}+\phi_{YY}+\phi_{ZZ} = 0 
\quad\mbox{for}\quad -H\left(X,Y\right)<Z<\eta\left(X,Y,t\right) \,,\\[2mm]
&&\displaystyle \phi_t+\fr\nabla\phi\cdot\nabla\phi+gZ = 0 \quad\mbox{on}\quad Z=\eta\left(X,Y,t\right) \,,\\[2mm]
&&\displaystyle \phi_Z = \eta_{t}+\phi_X\eta_X+\phi_Y\eta_Y \quad\mbox{on}\quad Z=\eta\left(X,Y,t\right) \,,\\[2mm]
&&\displaystyle \phi_Z+\phi_XH_X+\phi_YH_Y = 0 \quad\mbox{on}\quad Z=-H\left(X,Y\right) \,.
\end{array}
\right\}
\end{equation}
However, for the wave-structure interaction problem, the first component of the variational principle (\ref{Lagrangian-wave-body-fluid-interaction}) is coupled 
to the second component of (\ref{Lagrangian-wave-body-fluid-interaction}), and hence the variational Reynold's transport theorem 
(Flanders 1973; Daniliuk 1976; Gagarina, Van der Vegt \& Bokhove 2013) should be used 
for the variations $\delta\eta$ and $\delta\phi$, since the domain of integration $V\left(t\right)$ is time-dependent. 
Then, according to the usual procedure in the calculus of variations, 
and following the Euler-Poincar\'{e} variational framework introduced in Alemi Ardakani (2019), it can be proved that 
the variational principle (\ref{Lagrangian-wave-body-fluid-interaction}) for the variations 
$\delta\phi$, $\delta\eta$, $\delta\bm\Omega$, $\delta\bm Q$, $\delta\bm q$ and $\delta\dot{\bm q}$ 
subject to the restrictions that they vanish at the end points of the time interval, becomes 
\begin{equation}\nonumber
\left.
\begin{array}{rcl}
&&\displaystyle\delta\mathscr{L}\left(\phi,\eta,\bm\Omega,\bm Q,\bm q,\dot{\bm q}\right) = 
\int_{t_1}^{t_2}\int_{S_\eta} -\left(\phi_t+\fr\nabla\phi\cdot\nabla\phi
+gZ\right)\bigg|^{Z=\eta} \rho\,\delta\eta\,l^{-1} \rd S\,\rd t \\[4mm]
&&\displaystyle+\int_{t_1}^{t_2}\int_{S_w} 
\left(P\left(X,Y,Z,t\right)\bigg\langle\bm\Gamma,\bm x_w\times\bm n_b\bigg\rangle
+P\left(X,Y,Z,t\right)\bigg\langle\bm Q^{-1}\delta\bm q,\bm n_b\bigg\rangle\right) \rd S\,\rd t \\[4mm]
&&\displaystyle+\int_{t_1}^{t_2}\int_{S_\eta} 
\left(\eta_t+\eta_X\phi_X+\eta_Y\phi_Y-\phi_Z\right)\delta\phi\bigg|^{Z=\eta} \rho\,l^{-1} \rd S\,\rd t \\[4mm]
&&\displaystyle+\int_{t_1}^{t_2}\int_{S_w} 
\left(\dot{\bm X}_w\cdot\bm n-\frac{\partial\phi}{\partial \bm n}\right)\delta\phi \rho \,\rd S\,\rd t
+\int_{t_1}^{t_2}\int_{V\left(t\right)} \Delta\,\phi\,\delta\phi\,\rho \,\rd V\,\rd t 
\end{array}
\right\}
\end{equation}
\begin{equation}\label{wave-structure-interaction}
\left.
\begin{array}{rcl}
&&\displaystyle-\int_{t_1}^{t_2}\int_{S_b} \left(\phi_XH_X+\phi_YH_Y
+\phi_Z\right)\delta\phi\bigg|_{Z=-H} \rho \,\rd S\,\rd t \\[4mm]
&&\displaystyle+\int_{t_1}^{t_2}\int_{\mathcal{Q}} \bigg\langle\bm\Gamma,-\left(\bm x+\bm d\right)\times\left(\frac{D\bm u}{Dt}
+2\bm\Omega\times\bm u\right)\bigg\rangle \rho \,\rd\bm x\,\rd t \\[4mm]
&&\displaystyle+\int_{t_1}^{t_2}\bigg\langle\bm\Gamma,-m\overline{\bm x}\times\bm Q^{-1}\ddot{\bm q}
-\bm I_t\dot{\bm\Omega}-\bm\Omega\times\bm I_t\bm\Omega-mg\overline{\bm x}\times\bm\chi\bigg\rangle \,\rd t \\[4mm]
&&\displaystyle+\int_{t_1}^{t_2}\int_{\mathcal{Q}} \bigg\langle\bm Q^{-1}\delta\bm q,-\frac{D\bm u}{Dt}-2\bm\Omega\times\bm u
\bigg\rangle \rho \,\rd\bm x\,\rd t \\[4mm]
&&\displaystyle+\int_{t_1}^{t_2}\bigg\langle\bm Q^{-1}\delta\bm q,-m\bm Q^{-1}\ddot{\bm q}
-\dot{\bm\Omega}\times m\overline{\bm x}-\bm\Omega\times\left(\bm\Omega\times m\overline{\bm x}\right)
-mg\bm\chi\bigg\rangle \,\rd t = 0 \,,
\end{array}
\right\}
\end{equation}
where $\widehat{\bm\Gamma}=\bm Q^{-1}\delta\bm Q$ satisfies the so-called hat map, i.e. 
$\widehat{\bm\Gamma}\,\bm{\mathrm{r}}=\bm\Gamma\times\bm{\mathrm{r}}$ for any $\bm{\mathrm{r}}\in\mathbb{R}^3$ (Alemi Ardakani 2019), 
$\bm X_w$ denotes the position of a point on the wetted body surface $S_w$ relative to the 
spatial frame $\bm X$, $\bm n$ is the unit normal vector along $\partial V\supset S_w$ in the spatial frame $\bm X$, 
$l=\left(1+\eta_X^2+\eta_Y^2\right)^{1/2}$ giving $\rd S=l \,\rd X\rd Y$, $P$ is the pressure field of 
the exterior water waves defined by 
\begin{equation}\label{P}
P\left(X,Y,Z,t\right) = -\rho\left(\phi_t+\fr\nabla\phi\cdot\nabla\phi+gZ\right) 
\quad\mbox{on}\quad S_w \,,
\end{equation}
$\bm x_w$ is the position of a point on the wetted rigid body surface $S_w$ relative to the body frame $\bm x_b$, 
and $\bm n_b = \bm Q^{-1}\bm n$ is the unit normal vector along $S_w$ in the body frame $\bm x_b$. The derivation of 
the variational principle (\ref{wave-structure-interaction}) can be deduced from the variational derivations presented 
in Alemi Ardakani (2019). 
From (\ref{wave-structure-interaction}), we conclude that invariance of $\mathscr{L}$ with respect to a variation in the 
free-surface elevation $\eta$ yields the dynamic free-surface boundary condition in (\ref{exterior-equations}), 
invariance of $\mathscr{L}$ with respect to a variation in the velocity potential $\phi$ yields the 
field equation in (\ref{exterior-equations}) in the domain $V\left(t\right)$, 
invariance of $\mathscr{L}$ with respect to a variation in the 
velocity potential $\phi$ at $Z=-H\left(X,Y\right)$ gives the bottom boundary condition in (\ref{exterior-equations}), 
invariance of $\mathscr{L}$ with respect to a variation in the velocity potential $\phi$ at 
$Z=\eta\left(X,Y,t\right)$ gives the kinematic free-surface boundary condition in (\ref{exterior-equations}) and 
invariance of $\mathscr{L}$ with respect to a variation in the velocity potential $\phi$ on $S_w$ gives the 
contact condition on the wetted surface of the rigid body, 
\begin{equation}\label{contact-BC}
\displaystyle\frac{\partial\phi}{\partial \bm n} = \dot{\bm X}_w\cdot\bm n \quad\mbox{on}\quad S_w \,.
\end{equation}
Invariance of $\mathscr{L}$ with respect to $\bm\Gamma$ gives the hydrodynamic equation of motion for the 
rotational motion $\bm\Omega\left(t\right)$ of the floating rigid body interacting with the exterior water waves 
and dynamically coupled to its interior fluid motion 
\begin{equation}\label{Omega-equation-ship}
\left.
\begin{array}{rcl}
&&\displaystyle\int_{\mathcal{Q}} -\left(\bm x+\bm d\right)\times\left(\frac{D\bm u}{Dt}
+2\bm\Omega\times\bm u\right) \rho \,\rd\bm x-mg\overline{\bm x}\times\bm\chi
-m\overline{\bm x}\times\bm Q^{-1}\ddot{\bm q} \\[3mm]
&&\displaystyle-\bm I_t\dot{\bm\Omega}-\bm\Omega\times\bm I_t\bm\Omega
+\int_{S_w} P\left(X,Y,Z,t\right)\left(\bm x_w\times\bm n_b\right) \rd S = \bm 0 \,.
\end{array}
\right\}
\end{equation}
Invariance of $\mathscr{L}$ with respect to $\bm Q^{-1}\delta\bm q$ gives the hydrodynamic equation of motion for the 
translational motion $\bm q\left(t\right)$ of the floating rigid body containing fluid and in 
hydrodynamic interaction with the exterior water waves 
\begin{equation}\label{q-equation-ship}
\left.
\begin{array}{rcl}
&&\displaystyle\int_{\mathcal{Q}} \left(\frac{D\bm u}{Dt}+2\bm\Omega\times\bm u\right)
\rho \,\rd\bm x+m\bm Q^{-1}\ddot{\bm q}+\dot{\bm\Omega}\times m\overline{\bm x}
+\bm\Omega\times\left(\bm\Omega\times m\overline{\bm x}\right) \\[3mm]
&&\displaystyle+mg\bm\chi-\int_{S_w} P\left(X,Y,Z,t\right)\bm n_b \,\rd S = \bm 0 \,.
\end{array}
\right\}
\end{equation}
The terms including the pressure field $P\left(X,Y,Z,t\right)$ in the hydrodynamic equations of motion (\ref{Omega-equation-ship}) 
and (\ref{q-equation-ship}) are the moments and forces respectively acting on the rigid body due to interactions with the exterior water waves. 

The interior fluid of the floating rigid body satisfies the velocity potential 
theory of \S\ref{sec-Bernoulli-equation} and \S\ref{sec-Bateman-Luke-variational-principle}. Hence, after substituting 
for the velocity field $\bm u\left(\bm x,t\right)$ from (\ref{velocity-field}), the governing equations for the angular momentum 
(\ref{Omega-equation-ship}) and linear momentum (\ref{q-equation-ship}) of the floating rigid body 
dynamically coupled to its interior potential fluid sloshing while interacting with the exterior ocean waves become respectively 
\begin{equation}\label{Omega-equation-wave-structure-slosh}
\left.
\begin{array}{rcl}
&&\displaystyle\int_{\mathcal{Q}} -\left(\bm x+\bm d\right)
\times\left[\nabla\Phi_t+\nabla\left(\fr\nabla\Phi\cdot\nabla\Phi\right)
-\nabla\left(\nabla\Phi\cdot\left(\bm\Omega\times\left(\bm x+\bm d\right)+\bm Q^T\dot{\bm q}\right)\right)\right] 
\rho\,\rd\bm x \\[3mm]
&&\displaystyle-mg\overline{\bm x}\times\bm\chi
-m_v\overline{\bm x}_v\times\bm Q^{-1}\ddot{\bm q}-\bm I_v\dot{\bm\Omega}
+\bm I_v\bm\Omega\times\bm\Omega+\int_{S_w} P\left(X,Y,Z,t\right)\left(\bm x_w\times\bm n_b\right) \rd S = \bm 0 \,,
\end{array}
\right\}
\end{equation}
and
\begin{equation}\label{q-equation-wave-structure-slosh}
\left.
\begin{array}{rcl}
&&\displaystyle\int_{\mathcal{Q}} 
\left[\nabla\Phi_t+\nabla\left(\fr\nabla\Phi\cdot\nabla\Phi\right)
-\nabla\left(\nabla\Phi\cdot\left(\bm\Omega\times\left(\bm x+\bm d\right)+\bm Q^T\dot{\bm q}\right)\right)\right] 
\rho\,\rd\bm x \\[3mm]
&&\displaystyle+mg\bm\chi+m_v\bm Q^{-1}\ddot{\bm q}+\dot{\bm\Omega}\times m_v\overline{\bm x}_v
+\bm\Omega\times\left(\bm\Omega\times m_v\overline{\bm x}_v\right)-\int_{S_w} P\left(X,Y,Z,t\right)\bm n_b \,\rd S = \bm 0 \,.
\end{array}
\right\}
\end{equation}

In summary, the equations of motion for the exterior water waves in $V\left(t\right)$ are 
(\ref{exterior-equations}) with the contact boundary condition (\ref{contact-BC}). The equations of motion 
for the interior fluid of the rigid body are the field equation (\ref{Laplace-equation}) and the boundary conditions 
(\ref{no-flow-boundary-conditions-Phi-1}), (\ref{KFSBC-Phi}) and (\ref{DFSBC-Phi}), which are dynamically coupled to the hydrodynamic 
equations of motion for the floating rigid body (\ref{Omega-equation-wave-structure-slosh}) 
and (\ref{q-equation-wave-structure-slosh}). 

The tangent vectors $\dot{\bm Q}\left(t\right)\in \mbox{TSO}\left(3\right)$ along the integral 
curve in the rotation group $\bm Q\left(t\right)\in \mbox{SO}\left(3\right)$ may be retrieved via the reconstruction formula 
(Holm, Schmah \& Stoica 2009)
\begin{equation}\label{reconstruction-formula}
\dot{\bm Q} = \bm Q\widehat{\bm\Omega} \,. 
\end{equation}
The solution of (\ref{reconstruction-formula}) yields the integral curve 
$\bm Q\left(t\right)\in \mbox{SO}\left(3\right)$ for the orientation of the rigid body. 
Finally, diffirentiating the constraint equation $\bm\chi\left(t\right)=\bm Q^{-1}\left(t\right)\bm k$ gives 
\begin{equation}\label{chi-dot-equation}
\dot{\bm\chi}\left(t\right) = \bm\chi\left(t\right)\times\bm\Omega\left(t\right) \,\quad
\mbox{with}\quad \bm\chi\left(0\right)=\bm Q^{-1}\left(0\right)\bm k \,.
\end{equation}
So the evolutionary system for the rigid body motion (\ref{Omega-equation-wave-structure-slosh}) and (\ref{q-equation-wave-structure-slosh}) 
is completed by (\ref{reconstruction-formula}) and (\ref{chi-dot-equation}). 

\section{Concluding remarks}
\label{sec-concluding-remarks}

The paper is devoted to a new derivation of the Bernoulli equation for an inviscid and incompressible fluid sloshing 
in a container undergoing prescribed rigid-body motion in three dimensions. The Bernoulli equation is derived by integrating the Euler equations 
relative to the rotating and translating coordinate system attached to the moving container and using the vorticity equation. 
An alternative view on the Bateman-Luke variational principle is presented. It is shown that the Neumann boundary-value problem for the problem of 
potential fluid sloshing in a container undergoing 3--D rigid-body motion can be derived from the Bateman-Luke variational 
principle (\ref{pressure-integral}) with mathematically precise definitions of dependent and independent variables 
with respect to the spatial and body coordinate systems. A second variational principle is presented for 
the problem of 3--D interactions between potential water waves and a floating rigid body dynamically 
coupled to its interior potential fluid sloshing. The variational principle (\ref{Lagrangian-wave-body-fluid-interaction}) 
recovers the complete set of equations of motion for the exterior \emph{potential water waves} and the exact hydrodynamic equations of 
motion for the floating rigid body. The variational principle (\ref{Lagrangian-wave-body-fluid-interaction}) is coupled to 
the variational principle (\ref{pressure-integral}) which gives the full set of equations of motion for the interior \emph{potential 
fluid motion} of the rigid body. In Appendix \ref{sec-2D-VP}, the 3--D variational principles (\ref{pressure-integral}) and (\ref{Lagrangian-wave-body-fluid-interaction}) 
are reduced to two-dimensions to modify the variational principles given in Alemi Ardakani (2017). 

The presented variational principles (\ref{Lagrangian-wave-body-fluid-interaction}) and (\ref{pressure-integral}) 
and the corresponding partial differential equations for wave--structure--slosh interactions 
can be a starting point for further analytical and numerical analysis of 
dynamics of a liquid-filled spacecraft with interior potential fluid motion, potential fluid sloshing dynamics in moving tanks, 
a freely floating ship with fluid-filled tanks in hydrodynamic interaction with exterior water waves, and dynamics of 
floating structures such as ducted wave energy converters (Leybourne \emph{et al.} 2014). 
Gagarina \emph{et al.} (2014) developed a variational finite element method based on 
Luke's and Miles' variational principle (Luke 1967; Miles 1977) for nonlinear free surface gravity water waves. 
A direction of great interest is to extend the variational symplectic methods of Gagarina \emph{et al.} (2014, 2016) 
and Kalogirou \& Bokhove (2016) to develop hybrid numerical discretisations for the proposed Bateman-Luke variational principle (\ref{pressure-integral}) 
for the problem of potential water waves in rotating and translating coordinates, 
and the proposed variational principle (\ref{Lagrangian-wave-body-fluid-interaction}) 
for 3--D interactions between exterior surface waves and a floating structure with interior potential fluid sloshing. 

\vspace{3mm}
\begin{center}
\textbf{\Large Appendix}
\end{center}

\appendix 
\numberwithin{equation}{section}
\setcounter{equation}{0}
\section{Variational principles for two-dimensional interactions between potential water waves and a 
rigid body with interior potential fluid sloshing}
\label{sec-2D-VP}

The aim in this section is to reduce the proposed 3--D variational principles and their corresponding 
partial differential equations for the problem of two-dimensional interactions between potential surface waves and a 
floating rigid-body with interior potential fluid sloshing. 

\subsection{A variational principle for the interior fluid motion in two dimensions}
\label{sec-interior-fluid-2D}

For the 2-D problem, the fluid occupies the region $0\leq z \leq h\left(x,t\right)$ with $0\leq x \leq L$. 
The field equation (\ref{Laplace-equation}) for the interior fluid becomes
\begin{equation}\label{Laplace-equation-2D}
\Delta\,\Phi = \Phi_{xx}+\Phi_{zz} = 0 \quad\mbox{in}\quad \mathcal{Q}\left(t\right) \,.
\end{equation}
The rotation tensor $\bm Q\left(t\right)$ and the angular velocity vector $\bm\Omega$ take the form 
\begin{equation}\label{Q-2D}
\begin{array}{rcl}
\bm Q\left(t\right) = \left[
\begin{matrix} \hfill\cos\theta\left(t\right)\hfill & 0 & \hfill-\sin\theta\left(t\right) \\
0 & \hfill 1 & 0 \\
\hfill\sin\theta\left(t\right) & 0 & \hfill\cos\theta\left(t\right)
\end{matrix}
\right] \qand 
\bm\Omega = \left[
\begin{matrix}
0 \\
\hfill-\dot\theta \\
0
\end{matrix}
\right] \,.
\end{array} 
\end{equation}
The vessel is free to undergo pitch motion $\theta\left(t\right)$, surge motion $q_1\left(t\right)$ and heave motion $q_3\left(t\right)$. 

The velocity vector (\ref{velocity-field}) takes the form 
\begin{equation}\label{velocity-field-2D}
\begin{array}{rcl}
\bm u = \left(u,0,w\right)^T = \left[
\begin{matrix}
\Phi_x+\dot{\theta}\left(z+d_3\right)-\dot{q}_1\cos\theta-\dot{q}_3\sin\theta \\
0 \\
\Phi_z-\dot{\theta}\left(x+d_1\right)+\dot{q}_1\sin\theta-\dot{q}_3\cos\theta
\end{matrix}
\right] \,.
\end{array} 
\end{equation}
Relative to the body frame $\bm x$, the rigid-wall boundary conditions (\ref{no-flow-boundary-conditions-Phi-1}) are 
\begin{equation}\label{wall-bc-2D}
\left.
\begin{array}{rcl}
&&\displaystyle \Phi_x=-\dot{\theta}\left(z+d_3\right)+\dot{q}_1\cos\theta+\dot{q}_3\sin\theta \quad\mbox{at}\quad x=0 \quad\mbox{and}\quad x=L \,,\\[3mm]
&&\displaystyle \Phi_z=\dot{\theta}\left(x+d_1\right)-\dot{q}_1\sin\theta+\dot{q}_3\cos\theta \quad\mbox{at}\quad z = 0 \,.
\end{array}
\right\}
\end{equation}
The kinematic free surface boundary condition (\ref{KFSBC-Phi}) at $z=h\left(x,t\right)$ becomes 
\begin{equation}\label{KFSBC-2D}
\begin{array}{rcl}
&&\displaystyle\Phi_z-\dot{\theta}\left(x+d_1\right)+\dot{q}_1\sin\theta-\dot{q}_3\cos\theta = h_t
+\left(\Phi_x+\dot{\theta}\left(h+d_3\right)-\dot{q}_1\cos\theta-\dot{q}_3\sin\theta\right)h_x  \,,
\end{array}
\end{equation}
and the dynamic free surface boundary condition (\ref{DFSBC-Phi}) at $z=h\left(x,t\right)$ takes the form 
\begin{equation}\label{DFSBC-2D}
\left.
\begin{array}{rcl}
&&\displaystyle\Phi_t+\fr\left(\Phi_x^2+\Phi_z^2\right)
-\Phi_x\left(-\dot{\theta}\left(h+d_3\right)+\dot{q}_1\cos\theta+\dot{q}_3\sin\theta\right) \\[3mm]
&&\displaystyle-\Phi_z\left(\dot{\theta}\left(x+d_1\right)-\dot{q}_1\sin\theta+\dot{q}_3\cos\theta\right)
+g\sin\theta\left(x+d_1\right)+g\cos\theta\left(h+d_3\right) = 0 \,.
\end{array}
\right\}
\end{equation}

The Bateman-Luke variational principle (\ref{pressure-integral}) for the interior potential fluid sloshing in two dimensions becomes 
\begin{equation}\label{Bateman-Luke-2D}
\begin{cases}
\displaystyle\delta\mathscr{L}\left(\Phi,h\right) = \delta\int_{t_1}^{t_2}\int_0^{L}\int_0^h 
-\left[\Phi_t+\fr\nabla\Phi\cdot\nabla\Phi
-\Phi_x\left(-\dot{\theta}\left(z+d_3\right)+\dot{q}_1\cos\theta+\dot{q}_3\sin\theta\right) \right.\\[3mm]
\displaystyle\left.\hspace{1cm}-\Phi_z\left(\dot{\theta}\left(x+d_1\right)-\dot{q}_1\sin\theta+\dot{q}_3\cos\theta\right)
+g\sin\theta\left(x+d_1\right)+g\cos\theta\left(z+d_3\right)\right] 
\rho\,\rd z\,\rd x\,\rd t = 0 \,.
\end{cases}
\end{equation}
Now, according to the usual procedure in the calculus of variations, 
it can be proved that the variational principle (\ref{Bateman-Luke-2D}) becomes 
\begin{equation}\label{Bateman-Luke-2D-1}
\left.
\begin{array}{rcl}
&&\displaystyle\delta\mathscr{L}\left(\Phi,h\right) = \displaystyle
\int_{t_1}^{t_2}\int_0^L\left(h_t+h_x\Phi_x-\Phi_z
-h_x\left(-\dot\theta\left(h+d_3\right)+\dot{q}_1\cos\theta+\dot{q}_3\sin\theta\right) \right.\\[3mm]
&&\displaystyle\left.\hspace{3.7cm}+\dot\theta\left(x+d_1\right)
-\dot{q}_1\sin\theta+\dot{q}_3\cos\theta\right)\delta\Phi\bigg|^{z=h}\,\rho\,\rd x\,\rd t \\[3mm]
&&\displaystyle+\int_{t_1}^{t_2}\int_0^L\int_0^{h\left(x,t\right)}
\left(\Phi_{xx}+\Phi_{zz}\right)\delta\Phi \rho\, \rd z\,\rd x \,\rd t \\[3mm]
&&\displaystyle+\int_{t_1}^{t_2}\int_0^L\left(-\nabla\Phi\cdot\bm n_b
-\dot\theta\left(x+d_1\right)+\dot{q}_1\sin\theta-\dot{q}_3\cos\theta\right)
\delta\Phi\bigg|_{z=0}\rho\, \rd x\,\rd t \\[3mm]
&&\displaystyle+\int_{t_1}^{t_2}\int_0^{h\left(x,t\right)}
\left(-\nabla\Phi\cdot\bm n_b
+\dot\theta\left(z+d_3\right)-\dot{q}_1\cos\theta-\dot{q}_3\sin\theta\right)\delta\Phi\bigg|_{x=0}\rho\, \rd z\, \rd t \\[3mm]
&&\displaystyle+\int_{t_1}^{t_2}\int_0^{h\left(x,t\right)}\left(-\nabla\Phi\cdot\bm n_b
-\dot\theta\left(z+d_3\right)+\dot{q}_1\cos\theta+\dot{q}_3\sin\theta\right)\delta\Phi\bigg|_{x=L}\rho\, \rd z\, \rd t \\[3mm]
&&\displaystyle+\int_{t_1}^{t_2}\int_0^L p\left(x,h,t\right)\delta h \,\rd x\,\rd t = 0 \,,
\end{array}
\right\}
\end{equation}
where $\bm n_b=\bm Q^{-1}\bm n$ is outer normal to the boundary of $\mathcal{Q}\left(t\right)$ relative to the body frame $\bm x_b$ 
and $\bm n$ is the outer normal in the spatial frame $\bm X$. 
From (\ref{Bateman-Luke-2D-1}) it is obvious that invariance of $\mathscr{L}$ with respect to a 
variation in the free surface elevation $h$ yields the dynamic free surface boundary condition 
(\ref{DFSBC-2D}). Similarly, the invariance of $\mathscr{L}$ with respect to a variation in the velocity potential 
$\Phi$ yields the field equation (\ref{Laplace-equation-2D}). Also the invariance of $\mathscr{L}$ with respect to a 
variation in the velocity potential $\Phi$ at $z=0$, $x=0$, and $x=L$ recovers the 
rigid wall boundary conditions in (\ref{wall-bc-2D}). And the invariance of $\mathscr{L}$ with respect to 
a variation in the velocity potential $\Phi$ at $z=h\left(x,t\right)$ recovers the kinematic free surface boundary 
condition (\ref{KFSBC-2D}). 

\subsection{A variational principle for the exterior water waves and the motion of the rigid body containing potential fluid 
in two dimensions}
\label{sec-wave-ship-2D}

The 2--D variational principle (\ref{Bateman-Luke-2D}) gives the Neumann boundary-value problem for the interior 
fluid of the floating structure. The complete set of partial differential equations for the exterior surface waves 
and for the motion of the floating structure dynamically coupled to its interior potential fluid sloshing in the $\left(X,Z\right)$ plane 
can be obtained from the two-dimensional form of the variational principle (\ref{Lagrangian-wave-body-fluid-interaction}) by 
substituting (\ref{Q-2D}) and $\bm u=\left(u,0,w\right)$, $\bm q=\left(q_1,0,q_3\right)$, $\bm x=\left(x,0,z\right)$, $\bm d=\left(d_1,0,d_3\right)$, 
$\overline{\bm x}_v=\left(\overline{x}_v,0,\overline{z}_v\right)$ and 
$\fr\bm\Omega\cdot\bm I_v\bm\Omega=\fr m_v\left(\overline{x}_v^2+\overline{z}_v^2\right)\dot{\theta}^2$ 
into (\ref{Lagrangian-wave-body-fluid-interaction}). The 2--D variational principle takes the form 
\begin{equation}\label{Lagrangian-wave-body-fluid-interaction-2D}
\left.
\begin{array}{rcl}
&&\displaystyle\delta\mathscr{L}\left(\phi,\eta,\theta,q_1,q_3\right) 
= \displaystyle\delta\int_{t_1}^{t_2}\int_{V\left(t\right)}
-\rho\left(\phi_t+\fr\nabla\phi\cdot\nabla\phi+gZ\right) \rd V\,\rd t \\[4mm]
&&\displaystyle+\delta\int^{t_2}_{t_1} 
\int_0^L\int_0^{h\left(x,t\right)} \left(\fr\left(u^2+w^2\right)+\left(u-\dot\theta\left(z+d_3\right)\right)
\left(\dot q_1\cos\theta+\dot q_3\sin\theta\right) \right.\\[4mm]
&&\displaystyle\left. +\left(w+\dot\theta\left(x+d_1\right)\right)\left(-\dot q_1\sin\theta+\dot q_3\cos\theta\right)
+\dot\theta\left(w\left(x+d_1\right)-u\left(z+d_3\right)\right) \right.\\[4mm]
&&\displaystyle\left. +\fr\dot{\theta}^2\left(\left(x+d_1\right)^2+\left(z+d_3\right)^2\right)
+\fr\left(\dot q_1^2+\dot q_3^2\right)-g\sin\theta\left(x+d_1\right) \right.\\[4mm]
&&\displaystyle\left. -g\cos\theta\left(z+d_3\right)-gq_3\right) \rho\,\rd z\,\rd x\,\rd t \\[4mm]
&&\displaystyle+\delta\int_{t_1}^{t_2} \left(\fr m_v\left(\dot q_1^2+\dot q_3^2\right)
-m_v\overline{z}_v\dot\theta\left(\dot q_1\cos\theta+\dot q_3\sin\theta\right)
+\fr m_v\left(\overline{x}_v^2+\overline{z}_v^2\right)\dot{\theta}^2 \right.\\[4mm]
&&\displaystyle\left. +m_v\overline{x}_v\dot\theta\left(-\dot q_1\sin\theta+\dot q_3\cos\theta\right)
-m_vg\left(\overline{x}_v\sin\theta+\overline{z}_v\cos\theta+q_3\right)\right) \rd t = 0 \,,
\end{array}
\right\}
\end{equation}
where $\phi\left(X,Z,t\right)$ is the velocity potential of the exterior irrotational fluid. In the horizontal direction $X$, 
the fluid domain is cut off by a vertical surface $\mathcal{S}$ at $X=X_1, X_2$ which extends from the bottom to the free surface. 
The transient fluid domain $V\left(t\right)$ cosists of a fluid bounded by the impermeable bottom $S_b$ defined by the equation $Z=-H\left(X\right)$, 
the free surface $S_\eta$ defined by the equation $Z=\eta\left(X,t\right)$, the vertical surface $\mathcal{S}$ and the wetted surface 
$S_w$ of the rigid body interacting with the exterior water waves. It is proved in Appendix \ref{appendix-1} that taking the 
variations $\delta\phi$, $\delta\eta$, $\delta\theta$, $\delta q_1$ and $\delta q_3$ 
of the variational principle (\ref{Lagrangian-wave-body-fluid-interaction-2D}) 
gives the equations of motion for the exterior water waves in two-dimensions 
\begin{equation}\label{exterior-equations-2D}
\left.
\begin{array}{rcl}
&&\displaystyle \Delta\,\phi := \phi_{XX}+\phi_{ZZ} = 0 
\quad\mbox{for}\quad -H\left(X\right)<Z<\eta\left(X,t\right) \,,\\[3mm]
&&\displaystyle \phi_t+\fr\nabla\phi\cdot\nabla\phi+gZ = 0 \quad\mbox{on}\quad Z=\eta\left(X,t\right) \,,\\[3mm]
&&\displaystyle \phi_Z = \eta_t+\phi_X\eta_X \quad\mbox{on}\quad Z=\eta\left(X,t\right) \,,\\[3mm]
&&\displaystyle \phi_Z+\phi_XH_X = 0 \quad\mbox{on}\quad Z=-H\left(X\right) \,,
\end{array}
\right\}
\end{equation}
and the Euler-Lagrange equation for the pitch motion $\theta\left(t\right)$ of the floating rigid body  
\begin{equation}\label{theta-equation-2D}
\left.
\begin{array}{rcl}
&&\displaystyle\int_0^L\int_0^h \left(\frac{Dw}{Dt}\left(x+d_1\right)-\frac{Du}{Dt}\left(z+d_3\right)
+2\dot\theta\left(u\left(x+d_1\right)+w\left(z+d_3\right)\right)\right) \rho\,\rd z\,\rd x \\[4mm]
&&\displaystyle+m\overline{x}\left(-\ddot q_1\sin\theta+\ddot q_3\cos\theta\right)
-m\overline{z}\left(\ddot q_1\cos\theta+\ddot q_3\sin\theta\right)+mg\left(\overline{x}\cos\theta-\overline{z}\sin\theta\right) \\[3mm]
&&\displaystyle+I_t\ddot\theta+\int_{S_w} P\left(X,Z,t\right)\left(\bm x_w\times\bm n_b\right) \rd S = 0 \,,
\end{array}
\right\}
\end{equation}
where $P\left(X,Z,t\right)$ is defined in (\ref{pressure-2D}) and 
\begin{equation}\label{expressions-2D}
\left.
\begin{array}{rcl}
&&\displaystyle I_t=I_v+I_f=m_v\left(\overline{x}_v^2+\overline{z}_v^2\right)
+\int_0^L\int_0^h \left(\left(x+d_1\right)^2+\left(z+d_3\right)^2\right) \rho\,\rd z\,\rd x \\[4mm]
&&\displaystyle m\overline{x}=m_v\overline{x}_v+m_f\overline{x}_f=
m_v\overline{x}_v+\int_0^L\int_0^h \left(x+d_1\right) \rho\,\rd z\,\rd x \,,\quad m=m_v+m_f \,, \\[4mm]
&&\displaystyle m\overline{z}=m_v\overline{z}_v+m_f\overline{z}_f=
m_v\overline{z}_v+\int_0^L\int_0^h \left(z+d_3\right) \rho\,\rd z\,\rd x \,,
\end{array}
\right\}
\end{equation}
and the Euler-Lagrange equations for the translational motion of the floating rigid body in the 
surge $q_1\left(t\right)$ and heave $q_3\left(t\right)$ directions respectively 
\begin{equation}\label{q-1-2D}
\left.
\begin{array}{rcl}
&&\displaystyle\int_0^L\int_0^h \left(\frac{Du}{Dt}-2\dot\theta w\right) \rho\,\rd z\,\rd x
-m\overline{z}\,\ddot\theta-m\overline{x}\,\dot\theta^2
+m\left(\ddot q_1\cos\theta+\ddot q_3\sin\theta\right) \\[4mm]
&&\displaystyle+mg\sin\theta-\int_{S_w} P\left(X,Z,t\right)n_{1b} \,\rd S = 0 \,,
\end{array}
\right\}
\end{equation}
and 
\begin{equation}\label{q-3-2D}
\left.
\begin{array}{rcl}
&&\displaystyle\int_0^L\int_0^h \left(\frac{Dw}{Dt}+2\dot\theta u\right) \rho\,\rd z\,\rd x
+m\overline{x}\,\ddot\theta-m\overline{z}\,\dot\theta^2
-m\left(\ddot q_1\sin\theta-\ddot q_3\cos\theta\right) \\[4mm]
&&\displaystyle+mg\cos\theta-\int_{S_w} P\left(X,Z,t\right)n_{3b} \,\rd S = 0 \,.
\end{array}
\right\}
\end{equation}
Equations (\ref{theta-equation-2D}), (\ref{q-1-2D}) and (\ref{q-3-2D}) can be obtained from the 3--D Euler-Poincar\'{e} equations 
(\ref{Omega-equation-ship}) and (\ref{q-equation-ship}). 

Now, substitution of the velocity field (\ref{velocity-field-2D}) into the 
$\theta$-equation (\ref{theta-equation-2D}) gives 
\begin{equation}\label{theta-equation-2D-Phi}
\left.
\begin{array}{rcl}
&&\displaystyle\int_0^L\int_0^h \left(\dot\theta\left(x+d_1\right)\Phi_x+\dot\theta\left(z+d_3\right)\Phi_z
+\left(x+d_1\right)\Phi_{zt}-\left(z+d_3\right)\Phi_{xt} \right.\\[4mm]
&&\displaystyle\left.+\left(\Phi_{xz}\left(x+d_1\right)-\Phi_{xx}\left(z+d_3\right)\right)
\left(\Phi_x+\dot\theta\left(z+d_3\right)-\dot q_1\cos\theta-\dot q_3\sin\theta\right) \right.\\[4mm]
&&\displaystyle\left.+\left(\Phi_{zz}\left(x+d_1\right)-\Phi_{xz}\left(z+d_3\right)\right)
\left(\Phi_z-\dot\theta\left(x+d_1\right)+\dot q_1\sin\theta-\dot q_3\cos\theta\right)\right) \rho\,\rd z\,\rd x \\[4mm]
&&\displaystyle+I_v\ddot\theta-m_v\overline{z}_v\left(\ddot q_1\cos\theta+\ddot q_3\sin\theta\right)
+m_v\overline{x}_v\left(-\ddot q_1\sin\theta+\ddot q_3\cos\theta\right) \\[4mm]
&&\displaystyle+mg\left(\overline{x}\,\cos\theta-\overline{z}\,\sin\theta\right)
+\int_{S_w} P\left(X,Z,t\right)\left(\bm x_w\times\bm n_b\right) \rd S = 0 \,.
\end{array}
\right\}
\end{equation}
Substitution of the velocity field (\ref{velocity-field-2D}) into the 
$q_1$-equation (\ref{q-1-2D}) gives 
\begin{equation}\label{q1-equation-2D-Phi}
\begin{cases}
\displaystyle\int_0^L\int_0^h \left(\Phi_{xt}+\Phi_{xx}
\left(\Phi_x+\dot\theta\left(z+d_3\right)-\dot q_1\cos\theta-\dot q_3\sin\theta\right) \right.\\[2mm]
\displaystyle\left.\hspace{1.4cm}
+\Phi_{xz}\left(\Phi_z-\dot\theta\left(x+d_1\right)+\dot q_1\sin\theta-\dot q_3\cos\theta\right)
-\dot\theta\Phi_z\right) \rho\,\rd z\,\rd x \\[2mm]
\displaystyle-m_v\overline{z}_v\,\ddot\theta-m_v\overline{x}_v\,\dot\theta^2+m_v\left(\ddot q_1\cos\theta+\ddot q_3\sin\theta\right)
+mg\sin\theta-\int_{S_w} P\left(X,Z,t\right)n_{1b} \,\rd S = 0 \,.
\end{cases}
\end{equation}
Substitution of the velocity field (\ref{velocity-field-2D}) into the 
$q_3$-equation (\ref{q-3-2D}) gives 
\begin{equation}\label{q3-equation-2D-Phi}
\begin{cases}
\displaystyle\int_0^L\int_0^h \left(\Phi_{zt}+\Phi_{xz}
\left(\Phi_x+\dot\theta\left(z+d_3\right)-\dot q_1\cos\theta-\dot q_3\sin\theta\right) \right.\\[2mm]
\displaystyle\left.\hspace{1.4cm}
+\Phi_{zz}\left(\Phi_z-\dot\theta\left(x+d_1\right)+\dot q_1\sin\theta-\dot q_3\cos\theta\right)
+\dot\theta\Phi_x\right) \rho\,\rd z\,\rd x \\[2mm]
\displaystyle+m_v\overline{x}_v\,\ddot\theta-m_v\overline{z}_v\,\dot\theta^2-m_v\left(\ddot q_1\sin\theta-\ddot q_3\cos\theta\right)
+mg\cos\theta-\int_{S_w} P\left(X,Z,t\right)n_{3b} \,\rd S = 0 \,.
\end{cases}
\end{equation}
Equations (\ref{theta-equation-2D-Phi}), (\ref{q1-equation-2D-Phi}) and (\ref{q3-equation-2D-Phi}) can be obtained from 
the 3--D equations (\ref{Omega-equation-wave-structure-slosh}) and (\ref{q-equation-wave-structure-slosh}). 

\subsection{Proof of the Euler-Lagrange equations given in \S\ref{sec-wave-ship-2D}}
\label{appendix-1}

Applying the variational Reynold's transport theorem, the variational principle (\ref{Lagrangian-wave-body-fluid-interaction-2D}) 
for the variations $\delta\phi$, $\delta\eta$, $\delta\theta$, $\delta q_1$ and $\delta q_3$ becomes 
\begin{equation}\label{delta-Lagrangian-1}
\begin{cases}
\displaystyle\delta\mathscr{L}\left(\phi,\eta,\theta,q_1,q_3\right) = \int_{t_1}^{t_2}\int_{X_1}^{X_2} 
-\left(\phi_t+\fr\nabla\phi\cdot\nabla\phi+gZ\right)\bigg|^{Z=\eta} \rho\,\delta\eta\,\rd X\,\rd t \\[3mm]
\displaystyle+\int_{t_1}^{t_2}\int_{X_1}^{X_2} \left(\eta_t+\eta_X\phi_X-\phi_Z\right) 
\delta\phi\bigg|^{z=\eta} \rho\,\rd X\,\rd t
+\int_{t_1}^{t^2}\int_{V\left(t\right)} \varDelta\,\phi\,\delta\phi \,\rho\,\rd V\,\rd t \\[4mm]
\displaystyle+\int_{t_1}^{t_2}\int_{S_w} \left(\dot{\bm X}_w\cdot\bm n-\frac{\partial\phi}{\partial\bm n}\right) 
\delta\phi\,\rho\,\rd S\,\rd t
-\int_{t_1}^{t_2}\int_{X_1}^{X_2} \left(\phi_XH_x+\phi_Z\right)\delta\phi\bigg|_{Z=-H}\rho\,\rd X\,\rd t \\[4mm]
\displaystyle+\int_{t_1}^{t_2}\int_{S_w} P\left(X,Z,t\right) \left(\delta\bm X_w\cdot\bm n\right) \rd S\,\rd t \\[4mm]
\displaystyle+\int_{t_1}^{t_2}\int_0^L\int_0^h \left(-\delta\dot{\theta}\left(z+d_3\right)
\left(\dot q_1\cos\theta+\dot q_3\sin\theta\right)
+\delta\dot{\theta}\left(x+d_1\right)
\left(-\dot q_1\sin\theta+\dot q_3\cos\theta\right)\right.\\[3mm]
\displaystyle\left.+\left(u-\dot\theta\left(z+d_3\right)\right)
\left(\left(-\dot q_1\sin\theta+\dot q_3\cos\theta\right)\delta\theta
+\delta\dot q_1\cos\theta+\delta\dot q_3\sin\theta\right) \right.\\[3mm]
\displaystyle\left.+\left(w+\dot\theta\left(x+d_1\right)\right)
\left(\left(-\dot q_1\cos\theta-\dot q_3\sin\theta\right)\delta\theta
-\delta\dot q_1\sin\theta+\delta\dot q_3\cos\theta\right) \right.\\[3mm]
\displaystyle\left.+\delta\dot\theta\left(w\left(x+d_1\right)-u\left(z+d_3\right)\right)
+\delta\dot\theta\dot\theta\left(\left(x+d_1\right)^2+\left(z+d_3\right)^2\right)
+\delta\dot q_1\dot q_1+\delta\dot q_3\dot q_3\right.\\[3mm]
\displaystyle\left.-g\left(\cos\theta\left(x+d_1\right)-\sin\theta\left(z+d_3\right)\right)\delta\theta
-g\delta q_3 \right) \rho\,\rd Z\,\rd X\,\rd t \\[3mm]
\displaystyle+\int_{t_1}^{t_2} \left(m_v\left(\delta\dot q_1\dot q_1+\delta\dot q_3\dot q_3\right)
-m_v\overline{z}_v\delta\dot\theta\left(\dot q_1\cos\theta+\dot q_3\sin\theta\right)
+m_v\left(\overline{x}_v^2+\overline{z}_v^2\right)\delta\dot\theta\dot\theta \right.\\[2mm]
\displaystyle\left.-m_v\overline{z}_v\dot\theta\left(-\dot q_1\sin\theta+\dot q_3\cos\theta\right)\delta\theta
-m_v\overline{z}_v\dot\theta\left(\delta\dot q_1\cos\theta+\delta\dot q_3\sin\theta\right) \right.\\[2mm]
\displaystyle\left.+m_v\overline{x}_v\delta\dot\theta\left(-\dot q_1\sin\theta+\dot q_3\cos\theta\right)
+m_v\overline{x}_v\dot\theta\left(-\dot q_1\cos\theta-\dot q_3\sin\theta\right)\delta\theta \right.\\[2mm]
\displaystyle\left.+m_v\overline{x}_v\dot\theta\left(-\delta\dot q_1\sin\theta+\delta\dot q_3\cos\theta\right)
-m_vg\left(\overline{x}_v\cos\theta-\overline{z}_v\sin\theta\right)\delta\theta
-m_vg\delta q_3\right) \rd t = 0 \,,
\end{cases}
\end{equation}
where 
\begin{equation}\label{pressure-2D}
P\left(X,Z,t\right) = -\rho\left(\phi_t+\fr\nabla\phi\cdot\nabla\phi+gZ\right) \quad\mbox{on}\quad S_w \,,
\end{equation}
and it should noted that these variations are subject to the restrictions that they vanish at the end points of 
the time interval and on the vertical boundary at infinity $\mathcal{S}$. In (\ref{delta-Lagrangian-1}) 
$\bm X_w$ denotes the position of a point on the wetted vessel surface $S_w$ relative to 
the spatial frame $\bm X$, and $\bm n$ is the unit outward normal vector along $\partial V\supset S_w$ relative 
to the spatial frame $\bm X$. Note that 
\begin{equation}\label{variations-2D}
\left.
\begin{array}{rcl}
&&\displaystyle\int_{S_w} P\left(X,Z,t\right) \left(\delta\bm X_w\cdot\bm n\right) \rd S = 
\int_{S_w} P\left(X,Z,t\right) \left(\delta\bm Q\,\bm x_w+\delta\bm q\right)\cdot\bm n \,\rd S \\[3mm]
&&\displaystyle =\int_{S_w} P\left(X,Z,t\right) \left(\bm Q^\prime\bm x_w\,\delta\theta+\delta\bm q\right)
\cdot\bm Q\bm n_b \,\rd S \\[3mm]
&&\displaystyle =\int_{S_w}\left(-P\left(X,Z,t\right) \left(\bm x_w\times\bm n_b\right) \delta\theta
+P\left(X,Z,t\right) \left(\delta\bm q\cdot\bm Q\bm n_b\right)\right) \rd S \,,
\end{array}
\right\}
\end{equation}
where $\bm x_w$ is the position of a point on the wetted vessel surface relative to 
the body frame $\bm x_b$, $\bm n_b=\left(n_{1b},0,n_{3b}\right)$ is the unit outward normal vector along $\partial V\supset S_w$ relative 
to the body frame $\bm x_b$ and 
\begin{equation}\label{Q-prime}
\begin{array}{rcl}
&&\displaystyle{\bf Q}^\prime = 
\left[ \begin{matrix} \hfill-\sin\theta\hfill & \hfill0 & \hfill-\cos\theta\hfill \\
0 & \hfill0 & \hfill0\hfill \\
\hfill\cos\theta\hfill & \hfill0 & \hfill-\sin\theta\hfill \end{matrix} \,
\right] \,.
\end{array}
\end{equation}
Using the expression (\ref{variations-2D}), integrating by parts and applying the end point conditions, the variational principle 
(\ref{delta-Lagrangian-1}) simplifies to 
\begin{equation}\nonumber
\begin{cases}
\displaystyle\delta\mathscr{L}\left(\phi,\eta,\theta,q_1,q_3\right) = \int_{t_1}^{t_2}\int_{X_1}^{X_2} 
-\left(\phi_t+\fr\nabla\phi\cdot\nabla\phi+gZ\right)\bigg|^{Z=\eta} \rho\,\delta\eta\,\rd X\,\rd t \\[3mm]
\displaystyle+\int_{t_1}^{t_2}\int_{X_1}^{X_2} \left(\eta_t+\eta_X\phi_X-\phi_Z\right) 
\delta\phi\bigg|^{z=\eta} \rho\,\rd X\,\rd t
+\int_{t_1}^{t^2}\int_{V\left(t\right)} \Delta\,\phi\,\delta\phi \,\rho\,\rd V\,\rd t \\[3mm]
\displaystyle+\int_{t_1}^{t_2}\int_{S_w} \left(\dot{\bm X}_w\cdot\bm n-\frac{\partial\phi}{\partial\bm n}\right) 
\delta\phi\,\rho\,\rd S\,\rd t
-\int_{t_1}^{t_2}\int_{X_1}^{X_2} \left(\phi_XH_x+\phi_Z\right)\delta\phi\bigg|_{Z=-H}\rho\,\rd X\,\rd t \\[3mm]
\displaystyle+\int_{t_1}^{t_2}\int_{S_w}\left(-P\left(X,Z,t\right) \left(\bm x_w\times\bm n_b\right) \delta\theta
+P\left(X,Z,t\right) \left(\delta\bm q\cdot\bm Q\bm n_b\right)\right) \rd S\,\rd t \\[3mm]
\displaystyle+\int_{t_1}^{t_2}\int_0^L\int_0^h \left(-\frac{Dw}{Dt}\left(x+d_1\right)+\frac{Du}{Dt}\left(z+d_3\right)
+\left(z+d_3\right)\left(\ddot q_1\cos\theta+\ddot q_3\sin\theta\right) \right.\\[3mm]
\displaystyle\left.-\left(x+d_1\right)\left(-\ddot q_1\sin\theta+\ddot q_3\cos\theta\right)
-\ddot\theta\left(\left(x+d_1\right)^2+\left(z+d_3\right)^2\right) \right.\\[3mm]
\displaystyle\left.-2\dot\theta\left(u\left(x+d_1\right)+w\left(z+d_3\right)
-g\left(\cos\theta\left(x+d_1\right)-\sin\theta\left(z+d_3\right)\right)\right)\right) 
\delta\theta\,\rho\,\rd z\,\rd x\,\rd t \\[3mm]
\displaystyle+\int_{t_1}^{t_2}\int_0^L\int_0^h \left(\frac{Dw}{Dt}\sin\theta-\frac{Du}{Dt}\cos\theta
+2\dot\theta\left(u\sin\theta+w\cos\theta\right)-\ddot q_1 \right.\\[3mm]
\displaystyle\left.+\ddot\theta\left(\left(x+d_1\right)\sin\theta+\left(z+d_3\right)\cos\theta\right)
+\dot\theta^2\left(\left(x+d_1\right)\cos\theta-\left(z+d_3\right)\sin\theta\right)\right) 
\delta q_1\,\rho\,\rd z\,\rd x\,\rd t \\[3mm]
\displaystyle+\int_{t_1}^{t_2}\int_0^L\int_0^h \left(-\frac{Dw}{Dt}\cos\theta-\frac{Du}{Dt}\sin\theta
+2\dot\theta\left(w\sin\theta-u\cos\theta\right)-\ddot q_3-g \right.\\[3mm]
\displaystyle\left.+\ddot\theta\left(\left(z+d_3\right)\sin\theta-\left(x+d_1\right)\cos\theta\right)
+\dot\theta^2\left(\left(x+d_1\right)\sin\theta+\left(z+d_3\right)\cos\theta\right)\right) 
\delta q_3\,\rho\,\rd z\,\rd x\,\rd t \\[3mm]
\displaystyle+\int_{t_1}^{t_2} \left(-m_v\ddot q_1+m_v\overline{z}_v\left(\ddot\theta\cos\theta
-\dot\theta^2\sin\theta\right)+m_v\overline{x}_v\left(\ddot\theta\sin\theta+\dot\theta^2\cos\theta\right)
\right) \delta q_1\,\rd t 
\end{cases}
\end{equation}
\begin{equation}\label{delta-Lagrangian-2}
\begin{cases}
\displaystyle+\int_{t_1}^{t_2} \left(-m_v\ddot q_3+m_v\overline{z}_v\left(\ddot\theta\sin\theta
+\dot\theta^2\cos\theta\right)-m_v\overline{x}_v\left(\ddot\theta\cos\theta-\dot\theta^2\sin\theta\right)
-m_vg\right) \delta q_3\,\rd t \\[3mm]
\displaystyle+\int_{t_1}^{t_2} \left(m_v\overline{z}_v\left(\ddot q_1\cos\theta+\ddot q_3\sin\theta\right)
-m_v\overline{x}_v\left(-\ddot q_1\sin\theta+\ddot q_3\cos\theta\right)
-m_v\left(\overline{x}_v^2+\overline{z}_v^2\right)\ddot\theta \right.\\[3mm]
\displaystyle\left.-m_vg\left(\overline{x}_v\cos\theta-\overline{z}_v\sin\theta\right)\right) \delta\theta\,\rd t = 0 \,.
\end{cases}
\end{equation}
From (\ref{delta-Lagrangian-2}) we conclude that invariance of $\mathscr{L}$ with respect 
to a variation in the free surface elevation $\eta$ yields the dynamic free surface boundary condition 
in (\ref{exterior-equations-2D}), invariance of $\mathscr{L}$ with respect to a variation in 
the velocity potential $\phi$ yields the field equation in $V\left(t\right)$, 
invariance of $\mathscr{L}$ with respect to a variation in the velocity potential $\phi$ at $Z=-H\left(X\right)$ 
gives the bottom boundary condition in (\ref{exterior-equations-2D}), 
invariance of $\mathscr{L}$ with respect to a variation in the velocity potential $\phi$ at $Z=\eta\left(X,t\right)$ 
gives the kinematic free surface boundary condition in (\ref{exterior-equations-2D}), 
invariance of $\mathscr{L}$ with respect to a variation in the velocity potential $\phi$ on $S_w$ gives 
the contact condition on the vessel wetted surface 
\begin{equation}\label{contact-condition-2D}
\frac{\partial\phi}{\partial\bm n} = \dot{\bm X}_w\cdot\bm n \quad\mbox{on}\quad S_w \,.
\end{equation}
Finally, invariance of $\mathscr{L}$ with respect to $\delta\theta$ gives the Euler-Lagrange equation (\ref{theta-equation-2D}) for the 
rotational motion of the floating structure in the pitch direction $\theta\left(t\right)$, invariance of $\mathscr{L}$ with respect to $\delta q_1$ 
gives the Euler-Lagrange equation for the translational motion of the floating structure in the surge direction $q_1\left(t\right)$ 
\begin{equation}\label{surge-equation}
\left.
\begin{array}{rcl}
&&\displaystyle\int_0^{L}\int_0^h \left(\frac{Dw}{Dt}\sin\theta-\frac{Du}{Dt}\cos\theta
+\left(\left(x+d_1\right)\sin\theta+\left(z+d_3\right)\cos\theta\right)\ddot\theta \right.\\[2mm]
&&\displaystyle\left.+\left(\left(x+d_1\right)\cos\theta-\left(z+d_3\right)\sin\theta\right)\dot\theta^2
+2\dot\theta\left(u\sin\theta+w\cos\theta\right)-\ddot q_1\right) \rho\,\rd z\,\rd x \\[2mm]
&&\displaystyle-m_v\ddot q_1+m_v\left(\overline{x}_v\sin\theta+\overline{z}_v\cos\theta\right)\ddot\theta
+m_v\left(\overline{x}_v\cos\theta-\overline{z}_v\sin\theta\right)\dot\theta^2 \\[2mm]
&&\displaystyle+\int_{S_w} P\left(X,Z,t\right)\left(\cos\theta\,n_{1b}-\sin\theta\,n_{3b}\right) \rd S = 0 \,,
\end{array}
\right\}
\end{equation}
and invariance of $\mathscr{L}$ with respect to $\delta q_3$ 
gives the Euler-Lagrange equation for the translational motion of the floating structure in the heave direction $q_3\left(t\right)$ 
\begin{equation}\label{heave-equation}
\left.
\begin{array}{rcl}
&&\displaystyle\int_0^{L}\int_0^h \left(-\frac{Dw}{Dt}\cos\theta-\frac{Du}{Dt}\sin\theta
+\left(\left(z+d_3\right)\sin\theta-\left(x+d_1\right)\cos\theta\right)\ddot\theta \right.\\[3mm]
&&\displaystyle\left.+\left(\left(z+d_3\right)\cos\theta+\left(x+d_1\right)\sin\theta\right)\dot\theta^2
+2\dot\theta\left(w\sin\theta-u\cos\theta\right)-\ddot q_3-g\right) \rho\,\rd z\,\rd x \\[3mm]
&&\displaystyle-m_v\ddot q_3+m_v\left(\overline{z}_v\sin\theta-\overline{x}_v\cos\theta\right)\ddot\theta
+m_v\left(\overline{z}_v\cos\theta+\overline{x}_v\sin\theta\right)\dot\theta^2 \\[3mm]
&&\displaystyle-m_vg+\int_{S_w} P\left(X,Z,t\right)\left(\sin\theta\,n_{1b}+\cos\theta\,n_{3b}\right) \rd S = 0 \,.
\end{array}
\right\}
\end{equation}
Now, multiplying equations (\ref{surge-equation}) and (\ref{heave-equation}) by $\bm Q^T$ gives the Euler-Lagrange equations 
(\ref{q-1-2D}) and (\ref{q-3-2D}). 


\begin{thebibliography}{00}

\bibitem{haa-2017}
\textsc{Alemi Ardakani, H.} 2017 
A coupled variational principle for 2D interactions between water waves and a rigid body containing fluid. 
{\it J. Fluid Mech.} {\bf 827}, R2 1--12. 

\bibitem{haa-2019}
\textsc{Alemi Ardakani, H.} 2019 
A variational principle for three-dimensional interactions between water waves and a floating rigid body with 
interior fluid motion. 
{\it J. Fluid Mech.} {\bf 866}, 630--659.

\bibitem{haab-2011}
\textsc{Alemi Ardakani, H. \& Bridges, T. J.} 2011 
Shallow-water sloshing in vessels undergoing prescribed rigid-body motion in three dimensions. 
{\it J. Fluid Mech.} {\bf 667}, 474--519. 

\bibitem{bateman}
\textsc{Bateman, H.} 1932 
{\it Partial Differential Equations of Mathematical Physics.} 
Cambridge University Press, Cambridge.  

\bibitem{dgz}
\textsc{van Daalen, E. F. G., van Groesen, E. \& Zandbergen, P. J.} 1993 
A Hamiltonian formulation for nonlinear wave-body interactions. 
{\it In Proceedings of the Eight International Workshop on Water Waves and Floating Bodies, Canada} 
159--163. 

\bibitem{d-1976}
\textsc{Daniliuk, I. I.} 1976
On integral functionals with a variable domain of integration. In 
{\it Proceedings of the Steklov Institute of Mathematics}, vol. 118, pp. 1--44. American Mathematical Society.

\bibitem{frlt-2000}
\textsc{Faltinsen, O. M., Rognebakke, O. F., Lukovsky, I. A. \& Timokha, A. N.} 2000 
Multidimensional modal analysis of nonlinear sloshing in a rectangular tank with finite water depth. {\it J. Fluid Mech.} 
{\bf 407}, 201--234. 

\bibitem{ft-book}
\textsc{Faltinsen, O. M \& Timokha, A. N.} 2009 
{\it Sloshing.} 
Cambridge University Press, Cambridge. 

\bibitem{f-1973}
\textsc{Flanders, H.} 1973 
Differentiation under the integral sign. {\it Am. Math. Mon.} 
{\bf 80,} 615--627. 

\bibitem{ganvb}
\textsc{Gagarina, E., Ambati, V. R., Nurijanyan, S., van der Vegt, J. J. W. \& Bokhove, O.} 2016 
On variational and symplectic time integrators for Hamiltonian systems. {\it J. Comput. Phys.} 
{\bf 306,} 370--389. 

\bibitem{gavb}
\textsc{Gagarina, E., Ambati, V. R., van der Vegt, J. J. W. \& Bokhove, O.} 2014 
Variational space-time (dis)continuous Galerkin method for 
nonlinear free surface water waves. {\it J. Comput. Phys.} 
{\bf 275,} 459--483. 

\bibitem{gvb}
\textsc{Gagarina, E., van der Vegt, J. \& Bokhove, O.} 2013 
Horizontal circulation and jumps in Hamiltonian wave models. {\it Nonlinear Process. Geophys.} 
{\bf 20,} 483--500. 

\bibitem{gv-2003}
\textsc{Gerrits, J. \& Veldman, A. E. P.} 2003
Dynamics of liquid-filled spacecraft. {\it Journal of Engineering Mathematics} 
{\bf 45}, 21--38. 

\bibitem{kb}
\textsc{Kalogirou, A. \& Bokhove, O.} 2016 
Mathematical and numerical modelling of wave impact on wave-energy buoys. In 
{\it Proceedings of the International Conference on Offshore Mechanics and Arctic Engineering,} 
p. 8. The American Society of Mechanical Engineers. 

\bibitem{holm-book}
\textsc{Holm, D. D., Schmah, T. \&  Stoica, C.} 2009 
{\it Geometric Mechanics and Symmetry: From Finite to Infinite Dimensions.} 
Oxford University Press, Oxford. 

\bibitem{WEC-2014}
\textsc{Leybourne, M., Batten, W. M. J., Bahaj, A. S., Minns, N. \& O'Nians, J.} 2014 
Preliminary design of the OWEL wave energy converter pre-commercial demonstrator. {\it Renewable Energy} 
{\bf 61}, 51--56. 

\bibitem{luke}
\textsc{Luke, J. C.} 1967 
A variational principle for a fluid with a free surface. {\it J. Fluid Mech.} 
{\bf 27}, 395--397. 

\bibitem{l-1976}
\textsc{Lukovsky, I. A.} 1990 
{\it Introduction to Nonlinear Dynamics of a Solid Body with a Cavity including a Liquid.} 
Kiev: Naukova dumka (in Russian). 

\bibitem{l-2015}
\textsc{Lukovsky, I. A.} 2015  
{\it Nonlinear Dynamics: Mathematical Models for Rigid Bodies with a Liquid.} 
De Gruyter, Berlin. 

\bibitem{marsden-ratiu-book}
\textsc{Marsden, J. E. \&  Ratiu, T. S.} 1999 
{\it Introduction to Mechanics and Symmetry.} 
Springer-Verlag, New York. 

\bibitem{miles}
\textsc{Miles, J. W.} 1977 
On Hamilton's principle for surface waves. {\it J. Fluid Mech.} 
{\bf 83}, 153--158. 

\bibitem{mr-1968}
\textsc{Moiseyev, N. N. \& Rumyantsev, V. V.} 1968 
{\it Dynamic Stability of Bodies Containing Fluid.} 
Springer-Verlag, New York. 

\bibitem{t-2016}
\textsc{Timokha, A. N.} 2016 
The Bateman-Luke variational formalism in a sloshing with rotational flows. 
{\it Dopov. Nac. Akad. Nauk Ukr.} {\bf 4}, 30--34.  

\bibitem{vglhv-2007}
\textsc{Veldman, A. E. P., Gerrits, J., Luppes, R., Helder, J. A. \& Vreeburg, J. P. B.} 2007
The numerical simulation of liquid sloshing on board spacecraft. {\it J. Comput. Phys.} 
{\bf 224}, 82--99. 

\end{thebibliography}
\end{document}